\documentclass[preprint]{emulateapj}

\newcommand{\nustar}{{\it NuSTAR}}

\newcommand{\chandra}{{\it Chandra}}
\newcommand{\xmm}{{\it XMM-Newton}}
\newcommand{\suzaku}{{\it Suzaku}}
\newcommand{\xone}{IC~342 X-1}
\newcommand{\xtwo}{IC~342 X-2}

\def\lum{erg~s$^{-1}$}
\newcommand\msun  {M$_{\odot}$} 

\def\arcsec{$\,^{\prime\prime}$}

\def\arcmin{$\,^{\prime}$}

\def\simless{\mathbin{\lower 3pt\hbox
   {$\rlap{\raise 5pt\hbox{$\char'074$}}\mathchar"7218$}}} 
\def\simgreat{\mathbin{\lower 3pt\hbox
   {$\rlap{\raise 5pt\hbox{$\char'076$}}\mathchar"7218$}}} 
   
\def\ergflux{erg~cm$^{-2}$~s$^{-1}$}

\slugcomment{To be submitted to ApJ.}

\shorttitle{\xmm\ and \nustar\ spectroscopy of two ULXs in IC~342}
\shortauthors{Rana et al.}

\begin{document}


\title{The Broadband \xmm\ and \nustar\ X-ray Spectra of Two Ultraluminous X-ray Sources in the Galaxy IC~342}

\author{Vikram Rana\altaffilmark{1}, Fiona A. Harrison\altaffilmark{1},  
Matteo Bachetti\altaffilmark{2,3}, Dominic J. Walton\altaffilmark{1}, Felix Furst\altaffilmark{1}, Didier Barret\altaffilmark{2,3}, Jon M. Miller\altaffilmark{4}, Andrew C. Fabian\altaffilmark{5}, Steven E. Boggs\altaffilmark{6}, Finn C. Christensen\altaffilmark{7}, William W. Craig\altaffilmark{6,8}, Brian W. Grefenstette\altaffilmark{1}, Charles J. Hailey\altaffilmark{9}, Kristin K. Madsen\altaffilmark{1},  Andrew F. Ptak\altaffilmark{10}, Daniel Stern\altaffilmark{11}, Natalie A. Webb\altaffilmark{2,3}, William W. Zhang\altaffilmark{10}}
\altaffiltext{1}{Cahill Center for Astronomy and Astrophysics, California Institute of Technology, Pasadena, CA 91125}
\altaffiltext{2}{Universit\'e de Toulouse; UPS-OMP; IRAP; Toulouse, France}
\altaffiltext{3}{CNRS; Institut de Recherche en Astrophysique et
Plan\'etologie; 9 Av. colonel Roche, BP 44346, F-31028 Toulouse cedex 4, France}
\altaffiltext{4}{Department of Astronomy, University of Michigan, 500 Church Street, Ann Arbor, MI 48109-1042, USA}
\altaffiltext{5}{Institute of Astronomy, University of Cambridge, Madingley Road, Cambridge CB3 0HA, UK}
\altaffiltext{6}{Space Sciences Laboratory, University of California, Berke- ley, CA 94720, USA}
\altaffiltext{7}{DTU Space, National Space Institute, Technical University of Denmark, Elektrovej 327, DK-2800 Lyngby, Denmark}
\altaffiltext{8}{Lawrence Livermore National Laboratory, Livermore, CA 94550, USA}
\altaffiltext{9}{Columbia Astrophysics Laboratory, Columbia University, New York, NY 10027, USA}
\altaffiltext{10}{NASA Goddard Space Flight Center, Greenbelt, MD 20771, USA}
\altaffiltext{11}{Jet Propulsion Laboratory, California Institute of Technol- ogy, Pasadena, CA 91109, USA}

\begin{abstract}
We present results for two Ultraluminous X-ray Sources (ULXs), \xone\ and \xtwo, using two epochs of \xmm\ and \nustar\ observations separated by $\sim$7 days. We observe little spectral or flux variability above 1~keV between epochs, with unabsorbed 0.3--30~keV luminosities being $1.04^{+0.08}_{-0.06} \times 10^{40}$~erg~s$^{-1}$ for \xone\ and $7.40\pm0.20 \times 10^{39}$~erg~s$^{-1}$ for \xtwo, so that both were observed in a similar, luminous state. Both sources have a high absorbing column in excess of the Galactic value. Neither source has a spectrum consistent with a black hole binary in low/hard state, and both ULXs exhibit strong curvature in their broadband X-ray spectra. This curvature rules out models that invoke a simple reflection-dominated spectrum with a broadened iron line and no cutoff in the illuminating power-law continuum. X-ray spectrum of \xone\ can be characterized by a soft disk-like black body component at low energies and a cool, optically thick Comptonization continuum at high energies, but unique physical interpretation of the spectral components remains challenging. The broadband spectrum of \xtwo\ can be fit by either a hot (3.8~keV) accretion disk, or a Comptonized continuum with no indication of a seed photon population. Although the seed photon component may be masked by soft excess emission unlikely to be associated with the binary system, combined with the high absorption column, it is more plausible that the broadband X-ray emission arises from a simple thin blackbody disk component. Secure identification of the origin of the spectral components in these sources will likely require broadband spectral variability studies.
   
\end{abstract}


\section{Introduction}
\label{sec:intro}

Ultraluminous X-ray sources (ULXs), off-nuclear point X-ray sources with 
luminosities (assuming isotropic emission) exceeding $10^{39}$ \lum \citep{feng11} in the 0.3--10 keV band, are assumed to be black holes accreting from 
a binary companion.   A very small number of objects ($<6$) have X-ray luminosities in excess of $10^{41}$ \lum \citetext{\citealp{colbert99}, \citealp{farrell09}}.   For these
hyper-luminous X-ray sources \citetext{HLXs; \citealp{gao03}},  massive (10$^2$--10$^3$ M$_\odot$) black holes are the natural explanation, since relativistic beaming -- 
the only way to boost intrinsic luminosity by orders of magnitude -- has largely been ruled out by observations \citetext{\citealp{kording02}, \citealp{moon11}}. 
For the lower luminosity population, $L_x \lesssim 10^{41}$ \lum, high/super-Eddington accretion onto stellar mass (10--70 M$_\odot$) black holes with geometrical (i.e. non-relativistic) beaming  provide a reasonable explanation of the observed properties for many objects;  even with a moderately super-Eddington mass supply an apparent luminosity $\approx 20$ $L_{\rm{Edd}}$ can be reached \citep{ohsuga11}.   While most models focus on near- or super-Eddington accretion, we note that
recently one object with a luminosity exceeding $10^{40}$\lum\ has been shown to be an accreting neutron star \citep{bachetti14}.   

Most  `standard' ULXs 
have X-ray spectra that do not resemble the typical states of Galactic accreting black hole binaries 
\citetext{see \citealp{roberts07}, \citealp{feng11} for reviews}. Frequently
a soft excess, usually modeled as a disk blackbody
component with a temperature $\sim0.3$ keV is present, along with spectral curvature of continuum emission above $\sim$6-8~keV.  The spectral turnover is almost ubiquitous in high-quality X-ray spectra, and occurs at the upper end of the energy band accessible to \xmm, \chandra, and  \suzaku\ \citetext{\citealp{stobbart06}, \citealp{gladstone11}}.    In contrast, Galactic black hole binaries are predominantly found in a thin accretion disk-dominated soft state (effective black body temperature of $1 - 2$~keV) or a hard power-law state with spectral cutoffs seen only at very high energies ($>100$~keV).   

For standard ULXs  the 0.3--10~keV spectra can typically be decomposed into two components, the disk-like blackbody or thermal contribution prominent at energies $\simless1$~keV \citep{miller13}, and a harder component that can be modeled as  either a power law or a power law with an exponential cutoff, depending on the level of
spectral curvature seen above 6--8~keV  \citetext{\citealp{gladstone09}, \citealp{sutton13}}.   Towards the lower end of this luminosity range in some cases a single, broadened disk component (i.e. a disk with a shallower temperature profile than a standard Shakura \& Sunyaev thin disk) can dominate the broad band spectrum.   \citet{sutton13} classify the spectral states as `broadened disk'-dominated, predominantly seen at lower luminosity ($L_x \simless 0.3 \times 10^{40}$\lum), a `soft ultraluminous' state where the power law component has a steep spectral index ($\Gamma > 2$), and a `hard ultraluminous' state with $\Gamma < 2$ \citetext{see also \citealp{gladstone09}}.  The latter two spectral states are typically observed at higher luminosities, and
within this framework the difference is ascribed to viewing angle.

How to associate the different spectral components with specific physical regions in the accretion flow is unclear.   
The soft thermal component is often associated with an
accretion disk,  If the soft component is produced by a standard thin disk extending to the proximity of the black hole, it implies a black hole mass $M_{bh} \simgreat 100$\msun \citetext{\citealp{miller03, miller04}}.  The soft blackbody alternately plausibly originates in the photosphere of a thick, radiatively driven wind \citep{sutton13,king04}. Disk winds are expected for black holes close to Eddington, and become more prominent the higher the luminosity \citep{poutanen07}.   The trend for ULXs to exhibit higher variability in the soft ultra luminous state is consistent with the wind model, since the wind may become clumpy and result in time-variable obscuration \citep{sutton13}.  The wind scenario could explain the fact that the black body luminosity sometimes appears to be inversely correlated with temperature   \citetext{\citealp{kajava09}, \citealp{feng09}}, contrary to the  $L \propto T^4$ expected for accretion through an optically thick, geometrically thin disk. It should be noted, however, that while their sample did not include \xone\ and \xtwo,  \citet{miller13} found that a reanalysis of
data from a number of bright ULXs where an inverse correlation between luminosity and temperature has been claimed contradicts this finding. It has also been suggested
that the soft component arises from blurred line emission from highly ionized, fast-moving 
gas \citep{goncalves06}.

Several origins have also been suggested for the cutoff power law component.  
It can be modeled as thermal Comptonization from a cool ($\sim3$~keV), optically thick electron population 
associated with the inner regions of the accretion disk.   These electrons up scatter disk emission, creating the broad continuum with a turnover just below 10~keV.
This 'cool corona' may mask the hottest disk regions from view \citep{gladstone09}, explaining the low ($\sim0.3$~keV) temperatures
seen in the soft blackbody (in the scenario where the soft blackbody arises from the accretion disk).   Alternately, \citet{middleton11}
suggest the hard emission may be coming from the hot inner disk with the spectral shape resulting from a large color correction.  In the `broadened disk', or 'slim disk' scenario,
the broadband emission can be described by an advection-dominated disk with increased scale height 
\citep{abramowicz88} (i.e. a disk with a broadened temperature profile).  

The association of the observed spectral states of ULXs with stages of super-Eddington accretion is not, however definitive.  
Reflection models \citep{caballero-garcia10}
explain the $E > 6 - 8$ keV turnover as resulting from relativistically blurred Fe line emission reflected off the inner regions of a disk around an intermediate mass black hole.  
This model has been eliminated for NGC~1313 X-1 \citep{bachetti13}, Circinus ULX5 
\citep{walton13} and Holmberg IX X-1 \citep{walton14}, because \nustar\ fails to see the predicted upturn due to Compton scattering above 10~keV \citep{walton11}, however, it remains possible that some ULXs may harbor more massive black holes, and therefore exhibit accretion geometries distinct from those described above. 

The interpretation of the X-ray spectral components has been hampered by the limited 0.3--10~keV bandpass over which they have been studied.
In particular the spectral steepening above $\sim6$~keV could arise either from a cutoff or from poor modeling of the continuum due to possible broadened iron lines in a reflection-dominated regime 
\citetext{\citealp{caballero-garcia10}; \citealp{gladstone11}}.  Further, the shape of the turnover above 10~keV
can be used to better constrain physical models for the broad underlying continuum.
Unlike nearby bright Galactic binaries where collimated instruments can obtain quality broadband 
0.3--100~keV spectra, studies of ULXs above 10~keV require the sensitivity of a focusing telescope.   The {\em Nuclear Spectroscopic Telescope Array} \citetext{$NuSTAR$; \citealp{harrison13}} high energy X-ray focusing telescope, launched in 2012 June, is the first orbiting mission to provide sensitive spectroscopy in the 3--79~keV band.    \nustar\ is an ideal complement to \xmm, \chandra\ and \suzaku\ for spectral and temporal studies of ULXs.

In this paper we report results from observations of two ULXs in the face-on intermediate spiral galaxy IC~342 \citetext{$d = 3.93 \pm 0.10$~Mpc;  \citealp{tikhonov10}}
made as part of a joint \nustar\ and \xmm\ program to study a sample of nearby ($d \simless 10$~Mpc),  luminous (L$_x \sim 10^{40}$~\lum) ULXs.   \xone\ and \xtwo\ are ideally suited for broadband spectral studies with \nustar\ in that they are
bright, nearby and have relatively hard X-ray spectra.   Both are well-studied below 10~keV 
\citetext{\citealp{kubota02}, \citealp{feng09}}, and show long-term spectral variability.   State transitions similar to those seen in Galactic binaries have also been reported for both sources based on {\em ASCA} observations 
\citep{kubota01}, although source confusion due to the large PSF is a complicating factor for those data.

In \S~\ref{sec:observations} we summarize the observations and data reduction, \S~\ref{sec:spectra} describes the spectral modeling, \S~\ref{sec:timing} summarizes any timing variability during these
observations
and \S~\ref{sec:discussion} interprets the results in the context of previous X-ray observations. 
Finally, we present our conclusions in \S~\ref{sec:conclusions}.

\section{Observations and Data Reduction}
\label{sec:observations}

\begin{deluxetable*}{rrrrrrrrr}
\tablecolumns{7}
\tablewidth{0pc}
\tabletypesize{\scriptsize}
\tablecaption{\nustar\ observation log for IC 342. \label{nuobs}}
\tablehead{
\colhead{Obs ID} & \colhead{Start Time} & \colhead{End Time} &
\multicolumn{2}{c}{On time (ks) } &
\multicolumn{2}{c}{Source Counts\tablenotemark{*}} \\
\cline{4-5} \cline{6-7} \\
\colhead{ } & \colhead{UTC} & \colhead{UTC} &
\colhead{FPMA} & \colhead{FPMB} &
\colhead{FPMA} & \colhead{FPMB} }
\startdata
30002032003 & 2012-08-10 18:21:07 & 2012-08-12 13:51:07 & 98.6 & 98.6 & 4822 & 4638  \\
30002032005 & 2012-08-16 08:26:07 & 2012-08-18 19:16:07 & 127.7 & 127.7 & 7253 & 7110  \\
\enddata
\tablenotetext{*}{The counts are for on-axis source \xone.}
\end{deluxetable*}

\begin{deluxetable*}{rrrrrrrrr}
\tablecolumns{9}
\tablewidth{0pc}
\tabletypesize{\scriptsize}
\tablecaption{{\em XMM-Newton} observation log for IC 342. \label{xmmobs}}
\tablehead{
\colhead{Obs ID} & \colhead{Start Time} & \colhead{End Time} &
\multicolumn{3}{c}{On time (ks) } &
\multicolumn{3}{c}{Source Counts\tablenotemark{*}} \\
\cline{4-6} \cline{7-9} \\
\colhead{ } & \colhead{UTC} & \colhead{UTC} &
\colhead{pn} & \colhead{MOS1} & \colhead{MOS2} &
\colhead{pn} & \colhead{MOS1} & \colhead{MOS2} }
\startdata
0693850601 & 2012-08-11 20:07:29 & 2012-08-12 12:32:21 & 33.7 & 44.1 & 44.6 & 13843 & 6032 & 6119 \\
0693851301 & 2012-08-17 19:49:27 & 2012-08-18 11:51:48 & 33.0 & 39.1 & 39.1 & 19298 & 7609 & 7692 \\
\enddata
\tablenotetext{*}{The counts are for on-axis source \xone.}
\end{deluxetable*}

The \xmm\ and \nustar\ observations were performed as part of a joint program aimed at acquiring simultaneous
broad-band X-ray data on a sample of the brightest ULXs.   The \nustar\ observations were broken into two epochs
with integration times (corrected for Earth occultation and South Atlantic Anomaly (SAA) passage) of 98.6 and 127.7~ks,
respectively.   \xmm\ observed the field for overlapping intervals of 44 and 39~ks.  Tables~\ref{nuobs} and \ref{xmmobs} 
summarize the observations.

The \xmm\ data were reduced using  the Science Analysis System (SAS v12.0.1). 
We produced calibrated event files with EPCHAIN and EMCHAIN, created custom good time interval files to filter out periods of high background according to the prescription in the SAS manual, and selected only {\tt FLAG==0 \&\& PATTERN<4} events for EPIC-pn and {\tt FLAG==0 \&\& PATTERN<12} events for the EPIC-MOS cameras.  

We reduced the \nustar\  data using the \nustar\ 
Data Analysis Software (NuSTARDAS) v0.11.1 and CALDB version 20130509.  
The \nustar\ observations were taken during intervals of normal Solar activity,
and we used standard filtering to remove periods of high background during South Atlantic Anomaly (SAA) passages and
Earth occultation.   We created cleaned, calibrated event files using the {\tt NUPIPELINE} script with standard settings.   

For \xmm\  IC~342 X-1 was placed close to the optical axis.  Unfortunately, 
in the EPIC-pn camera IC~342 X-2 falls close to a chip gap, and the roll angle was such that the source was out of the field of view of
MOS1, and near a chip gap in MOS2.   For \xtwo\ we therefore report data only from EPIC-pn and MOS2.

\nustar\ \citep{harrison13} has two coaligned optics modules with corresponding focal planes, referred to as FPMA and FPMB.
In both \nustar\ modules \xone\ was placed on Detector~0, within 1.5\arcmin\ of the optical axis, and \xtwo\ was on Detector~2, neither falling near a gap \citetext{see \citealp{harrison13} for a description of the layout of the \nustar\ focal plane}.   As a result, \xtwo\ was about 6\arcmin\ off-axis where vignetting becomes significant above 20~keV \citetext{see \citealp{harrison13}}.

\begin{figure}[htbp] 
   \centering
  \includegraphics[width=\columnwidth]{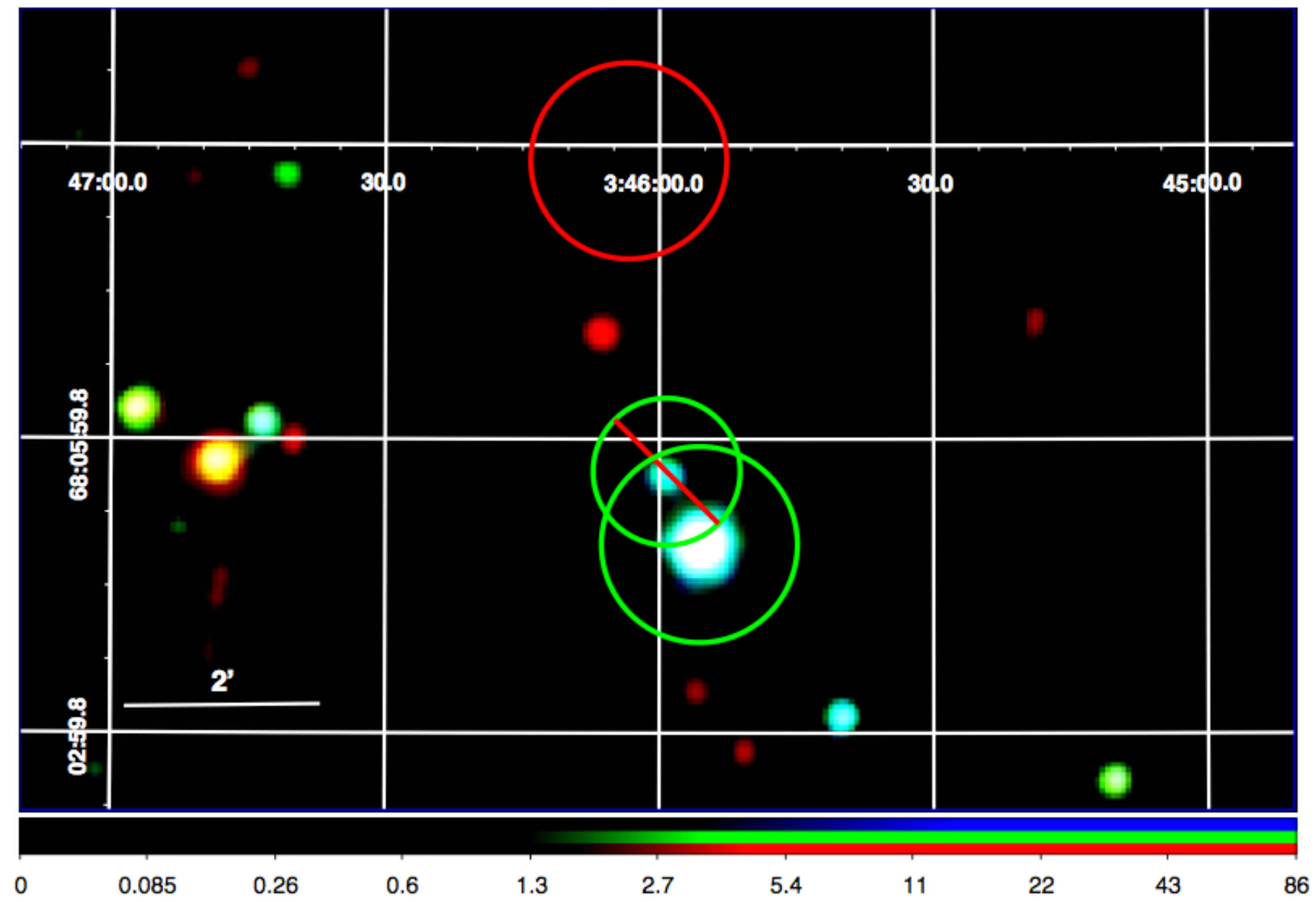} 
   \caption{\xmm\ field of view around \xone. Red, green and blue colors in the image indicate photons in the ranges  0.1--1.5, 1.5--4.5 and  4.5--12~keV, respectively. The larger green circle indicates the source extraction region, and the smaller green circle with the red line indicates the region excluded to remove a contaminating source.  The red circle is the background extraction region.}
   \label{fig:xmmextr}
\end{figure}

\begin{figure}[htbp] 
   \centering
  \includegraphics[width=\columnwidth]{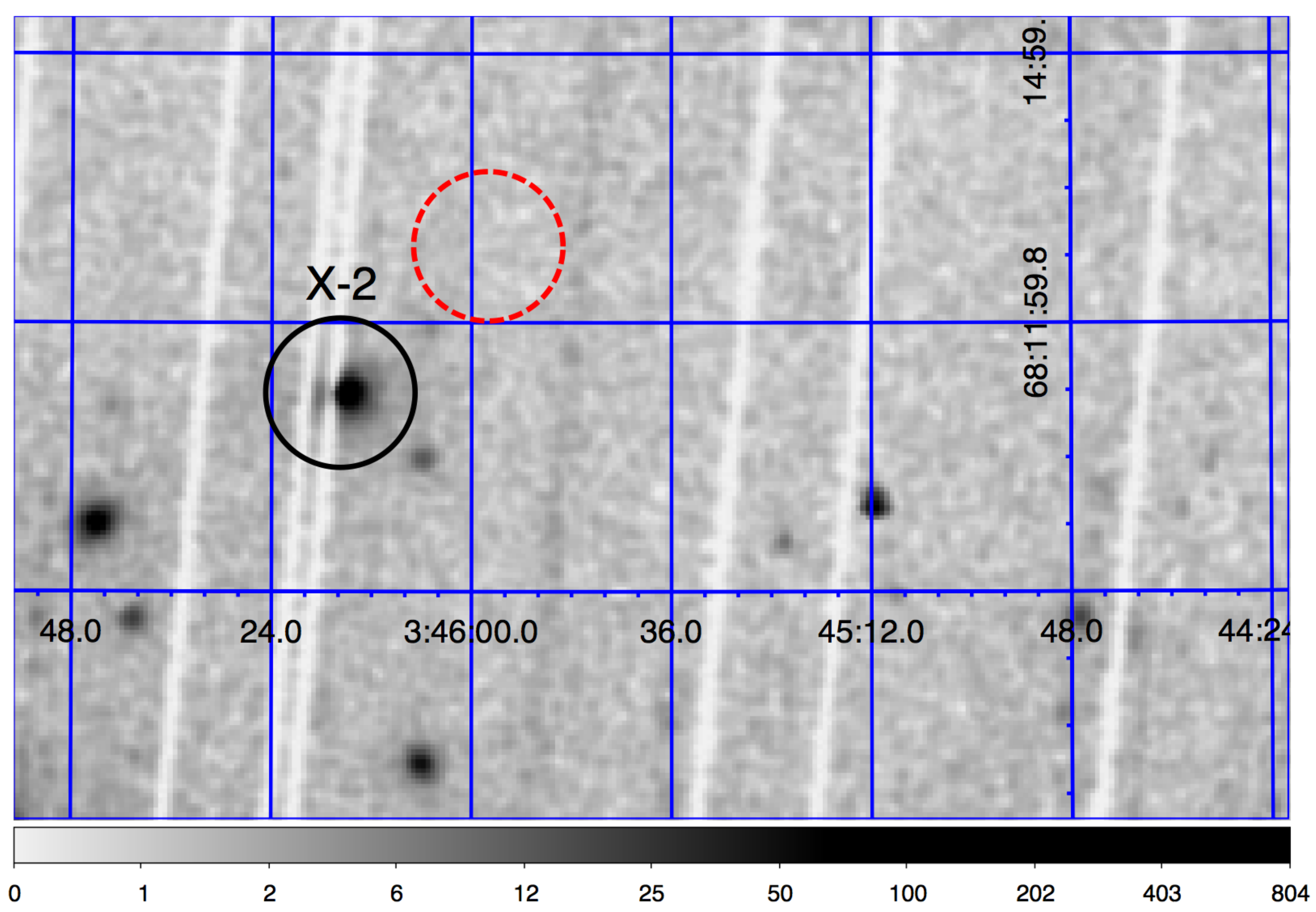} 
   \caption{\xmm\ pn field of view around \xtwo. Black circle shows source extraction region 
   and dashed red circle represents background region used for analysis. The source falls in the gap
   between two detectors.}
   \label{fig:xmmx2}
\end{figure}

\subsection{\xone}

To extract events for spectral analysis  we used a 60\arcsec\ radius region for \xmm\ and 50\arcsec\ for
\nustar.   A contaminating source near \xone\ is clearly visible in  a zoomed-in \xmm\  image (Figure~\ref{fig:xmmextr}).
In order to remove this object from the \xmm\ spectrum we exclude a 45\arcsec\ radius region centered on this
faint source.   Due to the larger \nustar\ PSF the contaminating source cannot be excised from the \nustar\ spectrum.  However,
the 5--10~keV flux of the contaminating source, as measured by \xmm,  is $8.1 \times 10^{-14}~$\ergflux, more than a factor 20 fainter than \xone. Using
the measured power law index of 1.7 to extrapolate to the \nustar\ band, we find that this source can contribute at
most 4\% of the total flux at $\sim$10~keV, but could contribute up to $\sim$20\% at 30 keV  if the contaminating source spectrum continues unbroken to this energy.
To verify that contamination does not affect our spectral results,
we compared \nustar\ spectra using a 30\arcsec\ extraction radius, and
confirmed the extracted spectra are entirely consistent with those obtained with the larger radius to within statistical errors.    It therefore appears that
the contaminating source has a spectral turnover below 30~keV.
We adopt the larger extraction radius for spectroscopic analysis in order to optimize the signal-to-noise.
To extract the background we used a 60\arcsec\ radius region to the NE of \xone\
for \xmm, and an 80\arcsec\ radius region on detector 0 for \nustar.
For EPIC-PN, it is prescribed by the manual that the background region should have a similar RAWY co-ordinates as the source (i.e., similar distance from the readout node), in order to have similar levels of low-energy noise, but in our case this was not feasible due to the presence of other sources and hot pixels at those co-ordinates. We made sure that the background region was extracted from the same detector, avoiding other sources, hot pixels and the detector column passing through the source\footnote{\href{http://xmm2.esac.esa.int/docs/documents/CAL-TN-0018.pdf}{http://xmm2.esac.esa.int/docs/documents/CAL-TN-0018.pdf}}. 
We used similar criteria for EPIC-MOS data also.

\begin{figure}
\includegraphics[width=\columnwidth]{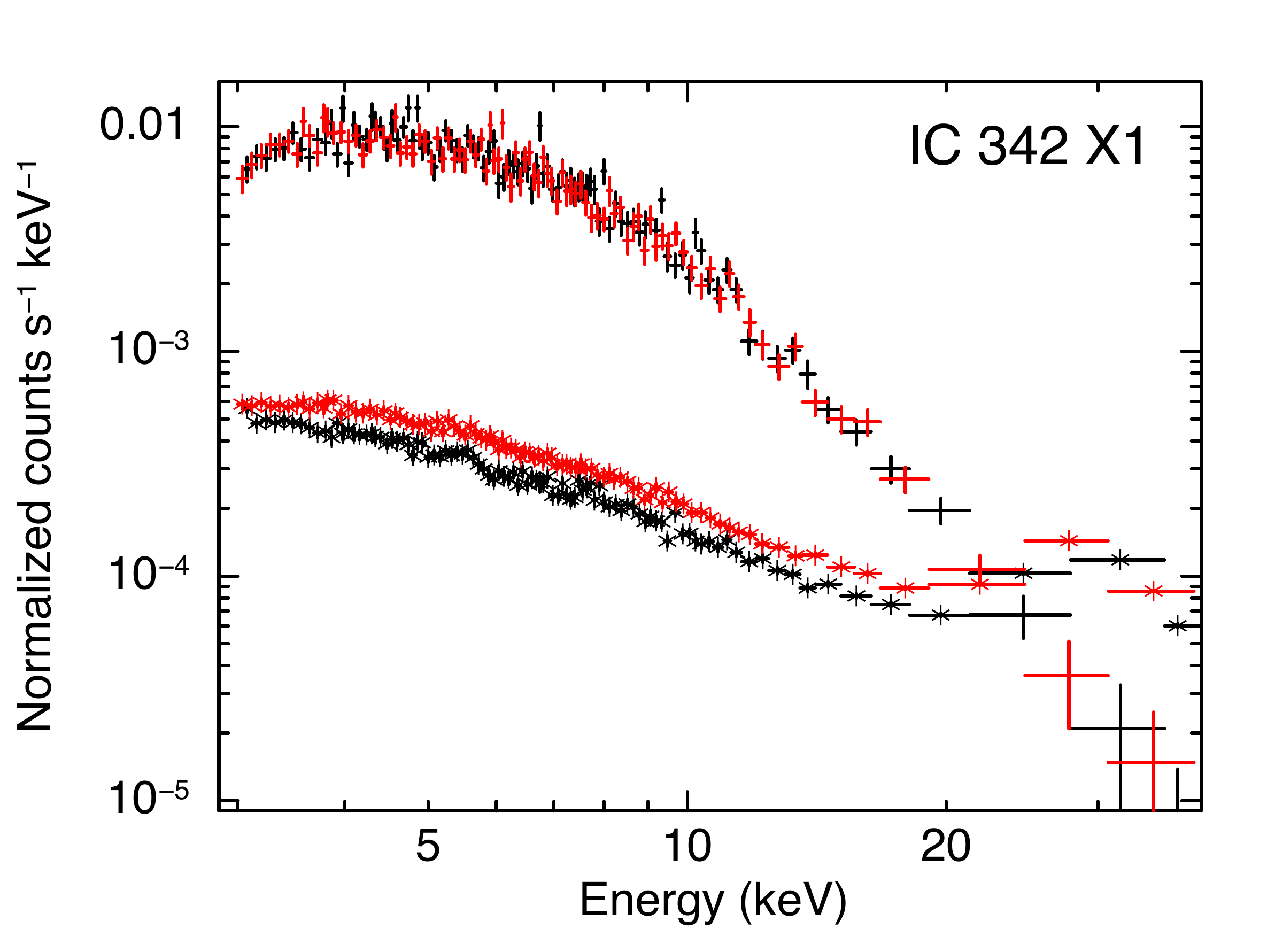}
\caption{\nustar\ count spectra (crosses) and background spectra (stars) extracted for \xone\ from the epoch 2 observation. Data from FPMA are shown in
black and from FPMB in red. Data have been rebinned for better visualization.}
\label{fig:x1raw}
\end{figure}

Figure~\ref{fig:x1raw} shows the  \nustar\ count spectrum from the epoch 2 observation
for both focal plane modules along with the extracted background spectrum.    
The source and background counts become comparable at approximately 25~keV, and spectra can be reliably analyzed
up to $\sim$30~keV.

\subsection{\xtwo}
The region surrounding \xtwo\ is clear of contaminating sources. For \xmm\ EPIC-pn and MOS2 we used a 50\arcsec\ circular region around the source and a region of the same size
for extracting background photons, chosen from a source free region on the same detector. We followed standard \xmm\ guidelines for generating spectral products.
For \nustar\ we extracted source counts
from a 50\arcsec\ region centered on the source position.   
For \nustar\ background estimation we used an 80\arcsec\ radius source free region on detector 2.

\begin{figure}
\includegraphics[width=\columnwidth]{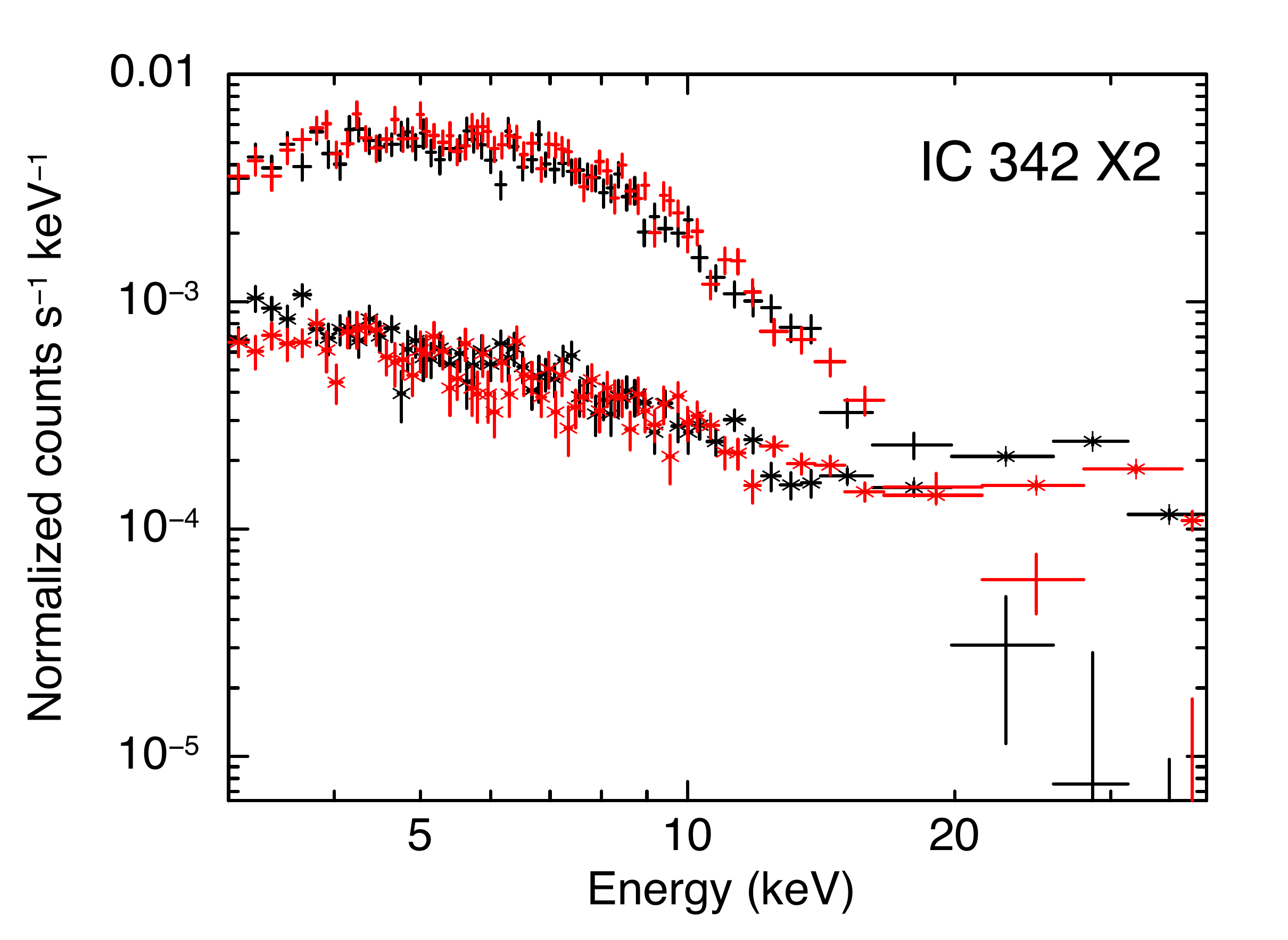}
\caption{\nustar\ count spectra (crosses) and background spectra (stars) extracted for \xtwo\ for the epoch 2 observation. Data from FPMA are shown in
black and from FPMB in red. Data have been rebinned for better visualization}
\label{fig:x2raw}
\end{figure}

Figure~\ref{fig:x2raw} shows the  \nustar\ count spectrum from the epoch 2 observation
for both focal plane modules along with the extracted background spectrum.    \xtwo\ is fainter than \xone, is further off-axis, and
the source and background counts become comparable just above 20~keV, so that reliable spectra can be analyzed up to $\sim$25~keV.
After making sure that the separate spectra for both the epochs do not show significant spectral 
variability, we combined the spectra from two epochs using ftool ADDASCASPEC. 

\section{Spectral Analysis}
\label{sec:spectra}

Throughout this work we perform spectral modeling using XSPEC v12.8.0 \citep{arnaud96}, and absorption by intervening
material is treated using {\tt tbabs} with updated  \citet{wilms00} solar abundances and photoionization 
cross-section as described in \citet{verner96}.  We perform fitting using $\chi^2$ minimization, and
quote all errors at 90\% confidence unless noted otherwise.   For fitting, we group spectra to a minimum of 25 counts per bin.

\begin{figure}
\includegraphics[width=\columnwidth]{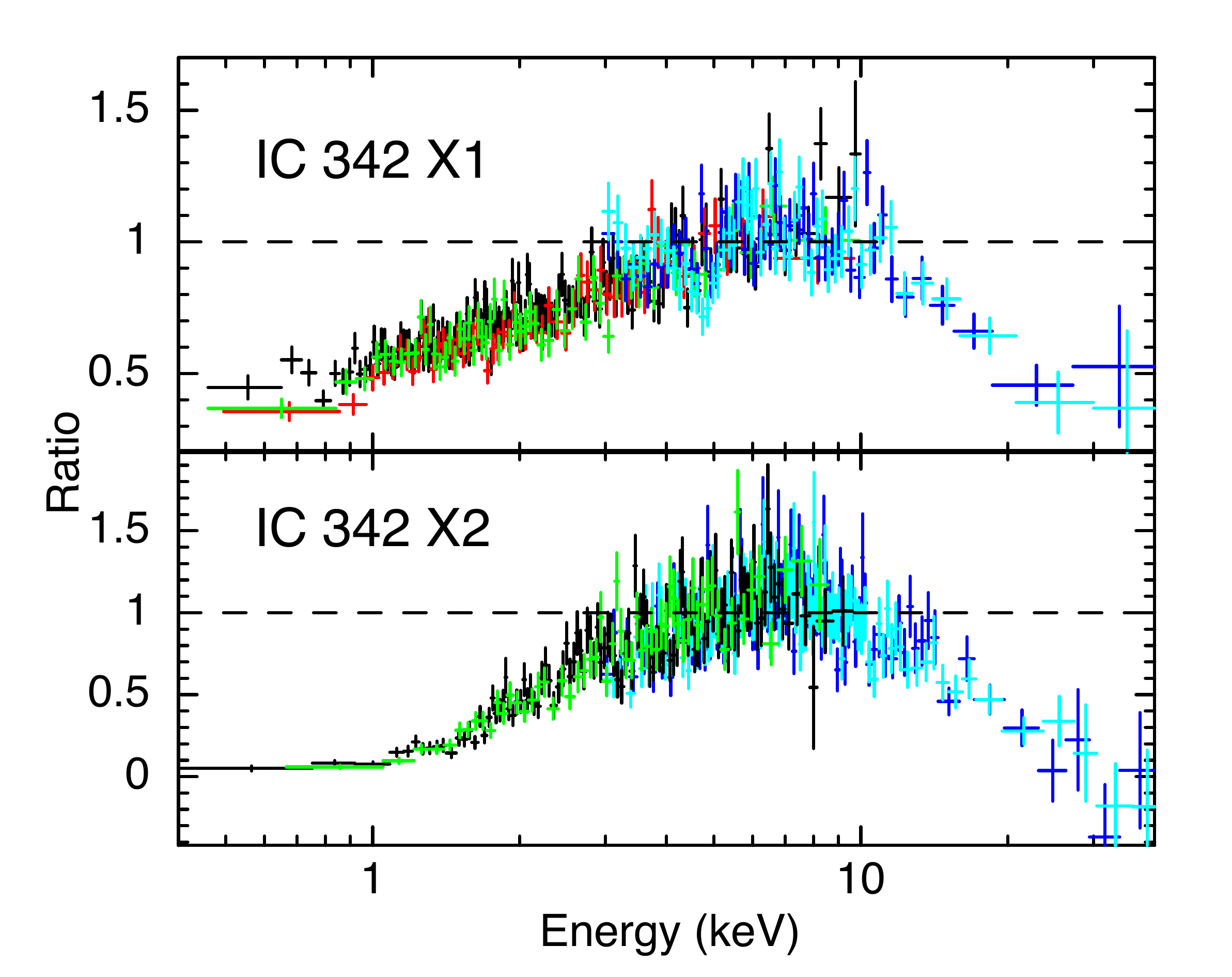}
\caption{Ratio of data to model for a power law fit in the overlapping 5--10~keV band.  \xmm\ pn is plotted in black, MOS1 in  red, MOS2
in green, \nustar\ FPMA in blue, and FPMB in light blue. The best-fit spectral index is $\Gamma = 2.22 \pm 0.10$ for \xone\ and
$\Gamma = 1.80 \pm 0.12$ for \xtwo. }
\label{fig-curvature}
\end{figure}

\begin{figure}
\includegraphics[width=\columnwidth]{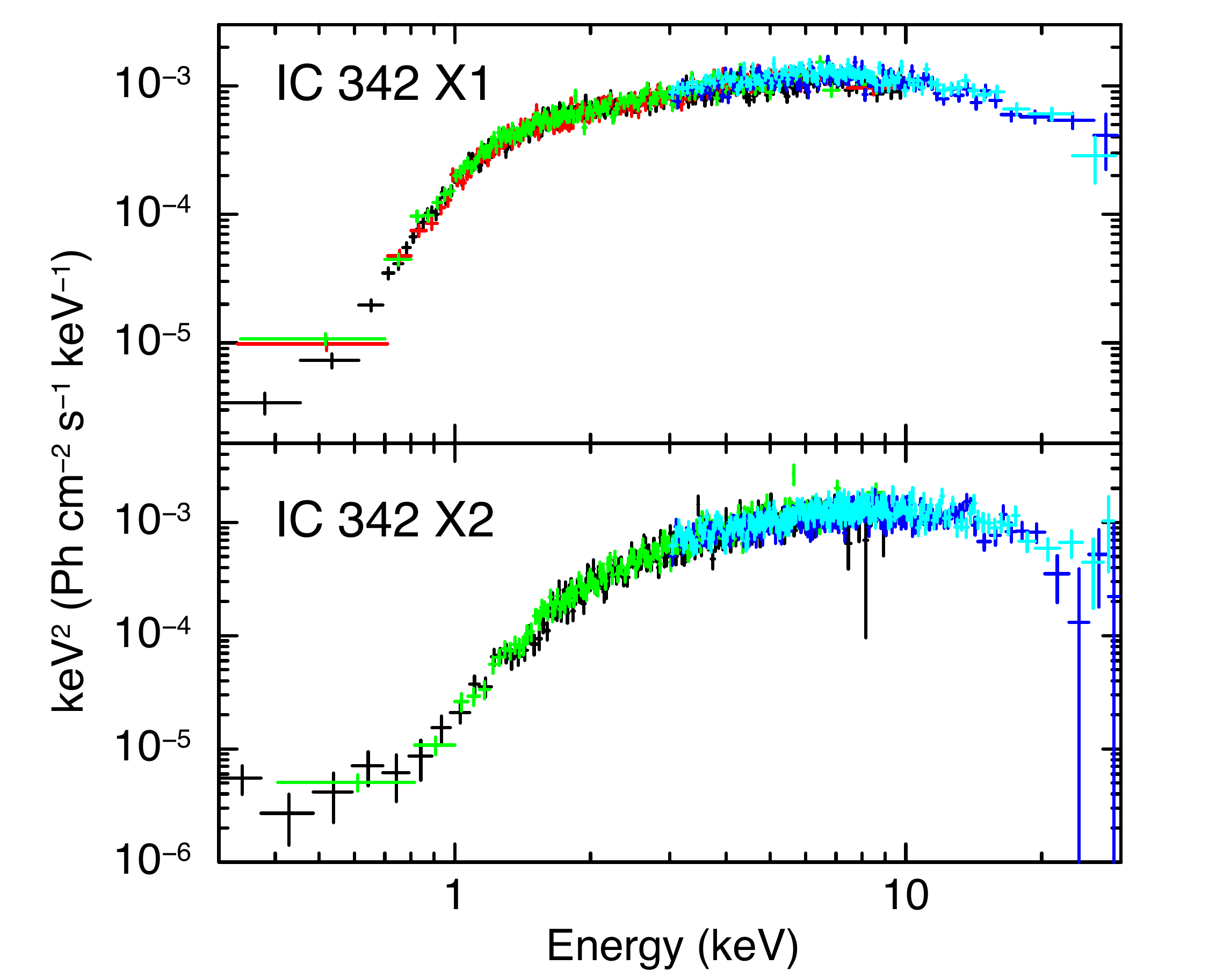}
\caption{The broad-band X-ray spectra of IC342 X-1 and X-2 unfolded through a model that simply consists of a constant.  \xmm\ pn is plotted in black, MOS1 in  red, MOS2
in green, \nustar\ FPMA in blue, and FPMB in light blue.}
\label{fig:spec_unfold}
\end{figure}

In order to assess the form of the broad-band \xmm\ and \nustar\ X-ray spectra of \xone\ and \xtwo, and specifically to look for spectral curvature above 6--8~keV observed previously in other ULXs
\citetext{\citealp{stobbart06}; \citealp{gladstone09}; \citealp{walton11}} we performed a joint fit of the data from
both epochs in the overlapping  5--10~keV energy band using a simple  power law model.   
We allowed the relative normalization between the \xmm\ EPIC-pn and MOS2, and \nustar\ FPMA and FPMB to vary in order to account for 
cross calibration uncertainties.  
For all models considered and for both sources we find the FPMA to FPMB cross calibration differences to be small ($<1$\%).  \xmm\ and \nustar\ have
significant overlap in their energy bands (both cover 3--10 keV), so that cross calibration factors are well-constrained.  The EPIC-pn to \nustar\ normalization differences are below the expected $\simless$10\% level.    
In Figure~\ref{fig-curvature} we plot the ratio of the data to the model over the 0.5--40~keV band.   The
continuum clearly has significant curvature across this broad band that is generally similar in both sources.   The best-fit 
5--10~keV spectral index is $\Gamma = 2.22 \pm 0.10$ for \xone\ and $\Gamma = 1.80 \pm 0.12$ for \xtwo. 
Figure~\ref{fig:spec_unfold} shows unfolded broad-band spectra through a simple constant model, for both sources. Although the spectra are broadly similar, it is evident that \xone\ exhibits a flatter spectrum from 1--4~keV, indicating an additional emission component at low energies.  In both sources the high-energy rollover is evident.

The luminosity of both sources was constant within 15\% between the two observations.  Using the best-fit {\tt diskbb + cutoffpl} model (see below), and using
a distance of 3.93~Mpc \citep{tikhonov10}, the absorbed 0.3--30~keV 
luminosity of \xone\ was $6.74^{+0.11}_{-0.37} \times 10^{39}$~erg~s$^{-1}$ during 
epoch 1 and  $7.56^{+0.07}_{-0.54} \times 10^{39}$~erg~s$^{-1}$ during epoch 2. 
For \xtwo\ the epoch 1 absorbed luminosity was $7.30^{+0.17}_{-0.46} \times 10^{39}$~erg~s$^{-1}$ compared to 
$6.83^{+0.15}_{-0.41} \times 10^{39}$~erg~s$^{-1}$ for epoch 2.   

\subsection{\xone\ Model Fits}

\begin{deluxetable*}{ccccc}
\tablecolumns{5}
\tablewidth{0pc}
\tabletypesize{\scriptsize}
\tablecaption{Best fit spectral parameters for \xone\ from joint fit to \xmm\
and \nustar\ data for various spectral models. \label{tab:fitsxone}}
\tablehead{
\colhead{Parameter} & \colhead{Unit} & \colhead{Epoch 1} & \colhead{Epoch 2} &
\colhead{Combined Epochs} }
\startdata
\cutinhead{Model = TBabs $*$ cutoffpl} \\
$N_H$ & 10$^{22}$ cm$^{-2}$ & 0.77$\pm$0.03 & 0.87$\pm$0.03 & 0.82$\pm$0.02 \\
$\Gamma$ &  & 1.45$\pm$0.06 & 1.61$\pm$0.05 & 1.55$\pm$0.04 \\
$E_{cut}$ & keV & 11.6$\pm$1.5 & 11.4$\pm$1.3 & 11.3$^{+0.9}_{-0.8}$ \\
$N_{cpl}$ & Photon keV$^{-1}$ cm$^{-2}$ s$^{-1}$  & (5.90$\pm$0.25)$\times$10$^{-4}$ & (8.83$\pm$0.37)$\times$10$^{-4}$ & (7.34$\pm$0.21)$\times$10$^{-4}$ \\
$\chi^2$/dof (Null hypo. Prob.) &  & 1499/1391 (0.02) & 1555/1592 (0.74) & 2509/2344 (9.1$\times10^{-3}$) \\
Flux & 10$^{-12}$ ergs cm$^{-2}$ s$^{-1}$ & 4.35$\pm$0.10 & 5.51$\pm$0.14 & 4.85$\pm$0.08 \\
Luminosity & 10$^{40}$ ergs s$^{-1}$ & 0.80$\pm$0.02 & 1.02$\pm$0.03 & 0.90$\pm$0.02 \\
\cutinhead{Model = TBabs $*$ diskbb}
$N_H$ & 10$^{22}$ cm$^{-2}$ & 0.40$\pm$0.014 & 0.41$\pm$0.013 & 0.41\tablenotemark{*} \\
$kT_{in}$ & keV & 2.45$\pm$0.05 & 2.23$\pm$0.04 & 2.31\tablenotemark{*} \\
$N_{bb}$ &  & (4.65$\pm$0.36)$\times$10$^{-3}$ & (7.90$\pm$0.54)$\times$10$^{-3}$ & 6.31$\times$10$^{-3}$\tablenotemark{*} \\
$\chi^2$/dof (Null hypo. Prob.) &  & 2462/1392 (2.0$\times 10^{-62}$) & 2972/1598 (1.6$\times10^{-85}$) & 4899/2345 (4.0$\times10^{-182}$) \\
Flux & 10$^{-12}$ ergs cm$^{-2}$ s$^{-1}$ & 3.51$\pm$0.07 & 4.09$\pm$0.07 & 3.76\tablenotemark{*} \\
Luminosity & 10$^{40}$ ergs s$^{-1}$ & 0.65$\pm$0.01 & 0.76$\pm$0.01 & 0.70\tablenotemark{*} \\
\cutinhead{Model = TBabs $*$ (diskbb + cutoffpl)}
$N_H$ & 10$^{22}$ cm$^{-2}$ & 1.02$^{+0.09}_{-0.08}$ & 1.01$^{+0.10}_{-0.08}$ & 1.02$^{+0.07}_{-0.06}$ \\
$kT_{in}$ & keV & 0.31$\pm$0.04 & 0.31$\pm$0.06 & 0.31$\pm$0.03 \\
$N_{bb}$ &  & 9.1$^{+10.4}_{-4.6}$ & 7.7$^{+18.3}_{-5.2}$ & 9.0$^{+8.0}_{-4.0}$ \\
$\Gamma$ &  & 1.00$\pm$0.13 & 1.39$\pm$0.12 & 1.22$\pm$0.09 \\
$E_{cut}$ & keV & 7.0$^{+1.0}_{-0.8}$ & 8.7$^{+1.3}_{-1.1}$ & 7.6$^{+0.8}_{-0.7}$ \\
$N_{cpl}$ & Photon keV$^{-1}$ cm$^{-2}$ s$^{-1}$ & (4.02$^{+0.53}_{-0.51}$)$\times$10$^{-4}$ & (7.37$\pm$0.85)$\times$10$^{-4}$ & (5.56$\pm$0.48)$\times$10$^{-4}$ \\
$\chi^2$/dof (Null hypo. prob.) &  & 1372/1389 (0.62) & 1518/1590 (0.90) & 2359/2342 (0.40) \\
Flux & 10$^{-12}$ ergs cm$^{-2}$ s$^{-1}$ & 5.21$^{+0.51}_{-0.36}$ & 6.10$^{+0.78}_{-0.47}$ & 5.64$^{+0.43}_{-0.32}$ \\
Luminosity & 10$^{40}$ ergs s$^{-1}$ & 0.98$^{+0.09}_{-0.07}$ & 1.11$^{+0.14}_{-0.09}$ & 1.04$^{+0.08}_{-0.06}$ \\
Flux Ratio\tablenotemark{**} &  & 0.40 & 0.21 & 0.32 \\
\cutinhead{Model = TBabs $*$ (diskbb + comptt)}
$N_H$ & 10$^{22}$ cm$^{-2}$ & 1.14$^{+0.13}_{-0.12}$ & 1.17$^{+0.15}_{-0.17}$ & 1.15$\pm$0.10 \\
$kT_{in}$ & keV & 0.22$\pm$0.03 & 0.18$\pm$0.02 & 0.20$\pm$0.02 \\
$N_{bb}$ &  & 62$^{+91}_{-37}$ & 172$^{+318}_{-128}$ & 97$^{+102}_{-52}$ \\
kT$_{e}$ & keV & 3.16$^{+0.20}_{-0.18}$ & 3.53$^{+0.27}_{-0.23}$ & 3.31$^{+0.16}_{-0.14}$ \\
$\tau$ &   & 6.33$\pm$0.38 & 5.13$\pm$0.33 & 5.65$\pm$0.25 \\
$N_{comptt}$ &  & (4.28$^{+0.43}_{-0.41}$)$\times$10$^{-4}$ & (6.58$\pm$0.72)$\times$10$^{-4}$ & (5.29$\pm$0.41)$\times$10$^{-4}$ \\
$\chi^2$/dof (Null hypo. prob.) &  & 1392/1389 (0.47) & 1537/1590 (0.83) & 2398/2342 (0.21) \\
Flux & 10$^{-12}$ ergs cm$^{-2}$ s$^{-1}$ & 6.07$^{+1.27}_{-0.83}$ & 7.62$^{+2.47}_{-1.69}$ & 6.83$^{+1.24}_{-0.92}$ \\
Luminosity & 10$^{40}$ ergs s$^{-1}$ & 1.12$^{+0.23}_{-0.15}$ & 1.41$^{+0.46}_{-0.31}$ & 1.26$^{+0.22}_{-0.17}$ \\
Flux Ratio\tablenotemark{**} &  & 0.45 & 0.40 & 0.42 \\ 
\cutinhead{Model = TBabs $*$ diskpbb}
$N_H$ & 10$^{22}$ cm$^{-2}$ & 0.81$\pm$0.03 & 0.92$\pm$0.03 & 0.87$\pm$0.02 \\
$kT_{in}$ & keV & 4.66$^{+0.29}_{-0.24}$ & 4.49$^{+0.25}_{-0.22}$ & 4.50$^{+0.18}_{-0.16}$ \\
p &  & 0.545$\pm$0.006 & 0.519$\pm$0.005 & 0.530$\pm$0.004 \\
$N_{bb}$ &  & (1.22$^{+0.35}_{-0.29}$)$\times$10$^{-4}$ & (1.25$^{+0.35}_{-0.29}$)$\times$10$^{-4}$ & (1.26$^{+0.24}_{-0.21}$)$\times$10$^{-4}$ \\
$\chi^2$/dof (Null hypo. prob.) &  & 1468/1391 (0.07) & 1547/1597 (0.81) & 2464/2344 (0.04) \\
Flux & 10$^{-12}$ ergs cm$^{-2}$ s$^{-1}$ & 4.54$\pm$0.10 & 5.86$\pm$0.14 & 5.13$\pm$0.08 \\
Luminosity & 10$^{40}$ ergs s$^{-1}$ & 0.84$\pm$0.02 & 1.08$\pm$0.03 & 0.95$\pm$0.02 \\
\cutinhead{Model = TBabs $*$ (powerlaw+kdblur2(reflionx)) }
$N_H$ & 10$^{22}$ cm$^{-2}$ & 1.09$\pm$0.02 & 1.08$\pm$0.03 & 1.06$\pm$0.02 \\
$\Gamma$ &  & 1.70$^{+0.02}_{-0.04}$ & 1.93$\pm$0.03 & 1.84$\pm$0.02 \\
$A_{Fe}$ &  & 5.7$^{+1.1}_{-0.8}$ & 3.1$^{+0.9}_{-0.8}$ & 4.9$^{+0.3}_{-0.4}$ \\
N$_{refl}$ &  & (1.32$^{+0.15}_{-0.04}$) $\times$ 10$^{-9}$ & (1.70$^{+0.08}_{-0.08}$) $\times$ 10$^{-9}$ & (1.55$^{+0.12}_{-0.03}$) $\times$ 10$^{-9}$ \\
R$_{in}$ & R$_g$ &  $<$1.33 & $<$1.25 & $<$1.24 \\
i & deg & 76.5$^{+2.5}_{-3.6}$ & 78.2$^{+2.0}_{-5.2}$ & 86.3$^{+0.2}_{-0.1}$ \\
$\chi^2$/dof (Null hypo. prob.) &  & 1563/1381 (4.1$\times10^{-4}$) & 1703/1590 (0.07) & 2747/2352 (1.5$\times10^{-11}$) \\
Flux & 10$^{-12}$ ergs cm$^{-2}$ s$^{-1}$ & 5.20$\pm$0.10 & 6.30$\pm$0.10 & 5.77$\pm$0.08 \\
Luminosity & 10$^{40}$ ergs s$^{-1}$ & 0.96$\pm$0.02 & 1.16$\pm$0.02 & 1.07$\pm$0.02 \\
\enddata
\tablenotetext{*}{The fit is unacceptable with poor reduced $\chi^2_\nu$=2.09, hence we did not estimate errors on the parameters.}
\tablenotetext{**}{The ratio of flux in soft to hard component in two component fit models.}
\tablecomments{All the listed flux and luminosity estimates are unabsorbed values in 0.3--30 keV energy band.}
\end{deluxetable*}

To characterize the broadband 0.3--30~keV continuum we fit the \xmm\ and \nustar\ data jointly with a number of parameterized models commonly used
to describe ULX X-ray spectra in the 0.3--10~keV band.   For simplicity in all fitting we apply a single neutral absorber to account for Galactic and local extinction, since the Galactic column is small ($N_H = 3.0 \times 10^{21}$~cm$^{-2}$) relative to that found in the spectral fitting.   
The absorption model ({\tt tbabs}) is an overall multiplicative factor for all models.  As noted above, a simple absorbed power law provides a poor description of the data, and so we focus on models, both empirical and physical, that have a cutoff at high energy.   

We fit six different continuum models that have a high energy turnover: 1) a power law with an exponential cutoff ({\tt cutoffpl} in XSPEC), 
2) a single blackbody disk fit to the entire spectrum,
3) a cutoff power law with the addition of a multicolor disk component with a temperature profile given by a \citet{shakura73} thin disk \citetext{{\tt diskbb}; \citealp{mitsuda84}},
4) a thin disk plus Comptonisation 
\citetext{{\tt comptt}; \citealp{titarchuk94}} model, where the seed photon temperature for
Compton scattering is set to the inner disk temperature,
5)  a disk model with a variable radial temperature profile index, $p$
\citep{mineshige94}, {\tt diskpbb}, and
6) a blurred relativistic reflection model that describes reflection of coronal emission from an accretion disk.   
The physical interpretation of the models and fit parameters will be discussed in \S~\ref{sec:discussion}; here we discuss the quality of the fits.

\begin{figure}
\includegraphics[width=\columnwidth]{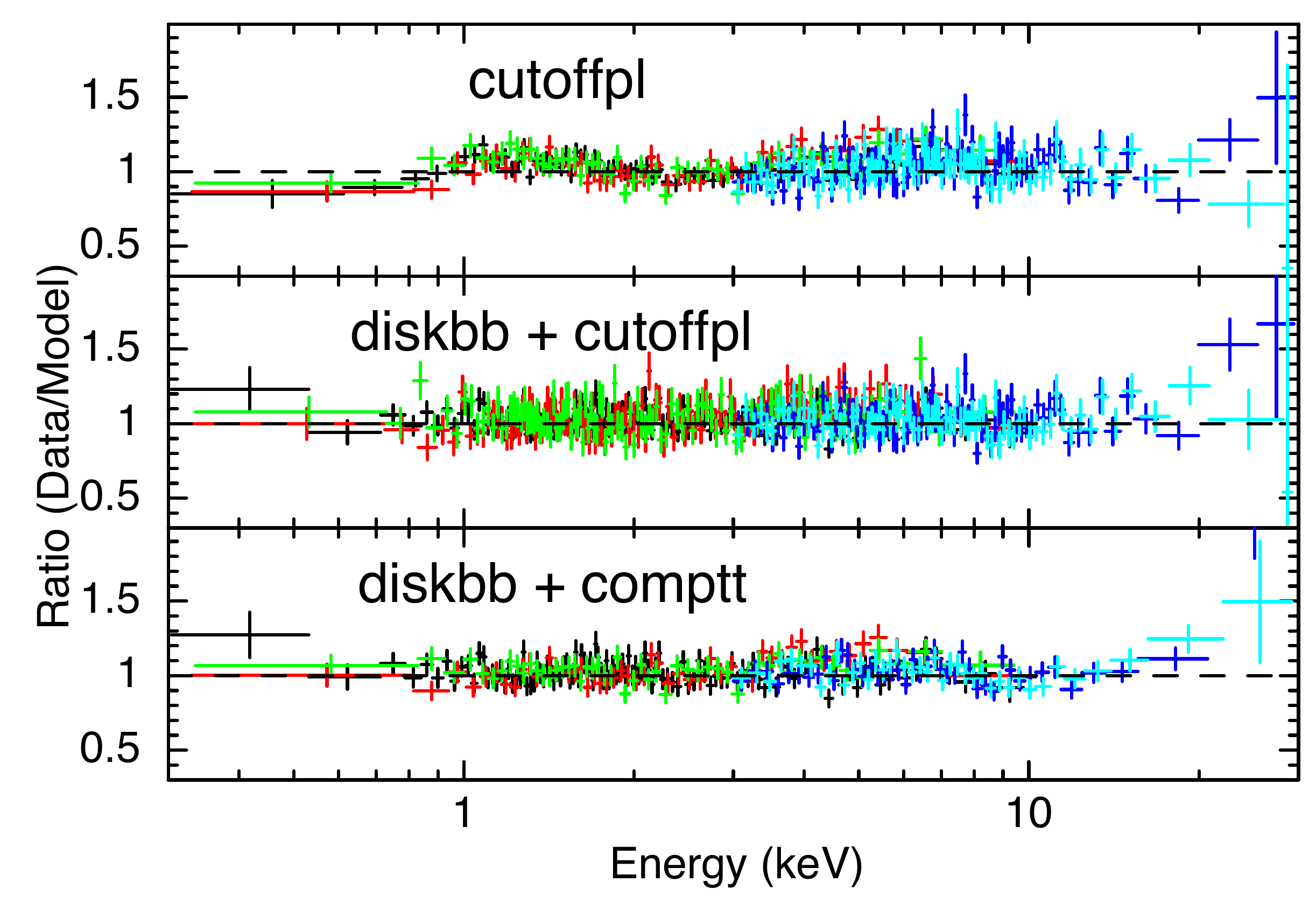}
\caption{Ratio of data to model for an absorbed power law with an exponential cutoff (top), an absorbed cutoff power law with a thin
accretion disk component (middle), and an accretion disk with a Comptonised continuum (bottom) for \xone.  \xmm\ pn is plotted in black, MOS1 in  red, MOS2
in green, \nustar\ FPMA in blue, and FPMB in light blue. Fit parameters for these models are given in Table~\ref{tab:fitsxone}.}
\label{fig:fitres1one}
\end{figure}

Table~\ref{tab:fitsxone} summarizes the best-fit parameters and goodness of fit for each model, along with the XSPEC model name. As listed in the table, we have used the set of all six models to characterize spectra from individual epochs as well as for the combined spectra. 
Although there are formally some differences in the best-fit parameters obtained from each epoch for the models that fit the broadband spectra well, these are very minor, and we mainly focus on describing the results from the combined data sets. For these models, we have also confirmed that fitting the data from each epoch simultaneously with all parameters linked between them and only overall normalization constants free to vary still provides excellent fits, justifying our decision to combine the data.

The empirical cutoff power law model provides a moderately acceptable fit, with $\chi^2_{\nu} = 1.07 (2509/2344)$ for the combined epoch 1 and 2 data.   However, the residuals show clear ``m"-shaped structure (see Figure~\ref{fig:fitres1one}) with an excess that is particularly evident at $\sim1$~keV.  
We therefore add a thermal disk component to the cutoff power law.  The best-fit inner temperature is
$0.31\pm0.03$~keV, and this significantly improves the fit, reducing the $\chi^2$ by 150 for two additional degrees of freedom and eliminating the systematic excess at low energy (see Figure~\ref{fig:fitres1one}).    We then investigated replacing the empirical cutoff power law with a Comptonization component with the seed photon energy tied to the the inner blackbody disk temperature.   This also results in an acceptable fit ($\chi^2_{\nu} = 1.02 (2398/2342)$), although the residuals for the three spectral bins above $\sim15$~keV are systematically high. We have also investigated disconnecting seed photon temperature from inner disk temperature while fitting absorbed $diskbb + comptt$ models. However, this leads to a strong degeneracy between the two parameters, making one of them completely unconstrained without affecting the fit statistically. Therefore, we decided to proceed with the two parameters tied together for this model fit, following the limitations of the present data.

\begin{figure}
\includegraphics[width=\columnwidth]{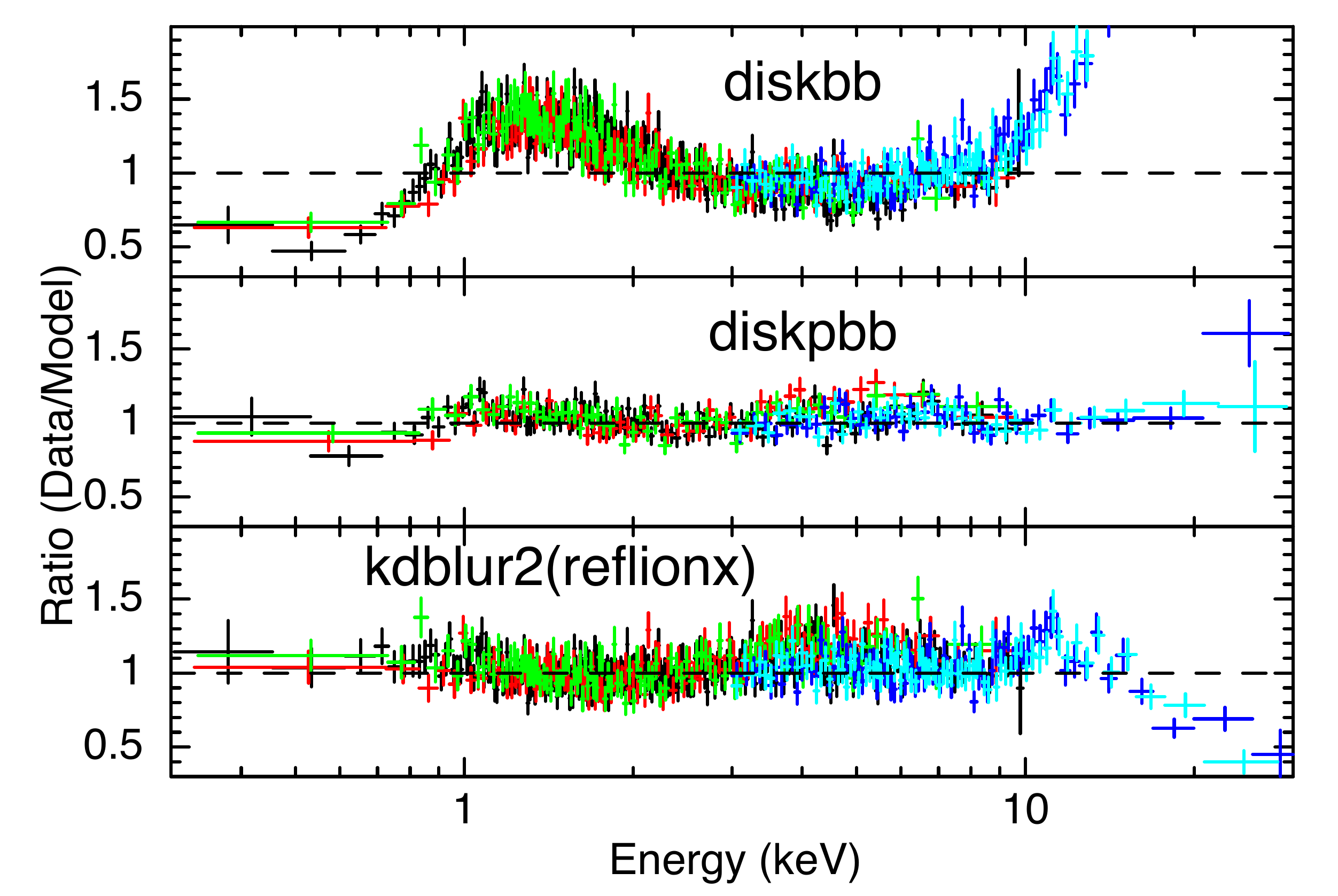}
\caption{Ratio of data to model for a Shakura \& Sunyaev thin disk (top),  an accretion disk with advection (middle), and a relativistic blurred reflection model (bottom) for \xone.   \xmm\ pn is plotted in black, MOS1 in  red, MOS2
in green, \nustar\ FPMA in blue, and FPMB in light blue.  Fit parameters for these models are given in Table~\ref{tab:fitsxone}.}
\label{fig:fitres1two}
\end{figure}

We also attempted to fit an absorbed {\tt diskbb} model alone to the data.  Note this is different from the previous cases where we add the disk component to explain a soft excess on top of the broad continuum.  We find a hot Shakura \& Sunyaev thin disk cannot explain the observed broadband spectrum ($\chi^2_{\nu} = 2.09 (4899/2345)$; see also Figure~\ref{fig:fitres1two}).    At accretion rates that reach a significant fraction of Eddington the scale height of the disk is expected to increase, and advection becomes important \citep{abramowicz88} resulting in a shallower radial temperature profile and broader spectral emission relative to a thin disk.  We therefore tried fitting the data with the {\tt diskpbb} model, which allows the radial temperature profile index, $p$ to be a free parameter.    This model is a significantly better fit compared to the thin disk,  yielding an acceptable fit ($\chi^2_{\nu} = 1.05 (2464/2344)$).   The best-fit value for $p$ is 
$0.53\pm0.004$ (constrained to better then 1\%), shallower than the $p=0.75$ associated with a thin disk.

For comparison, we also fit the 0.3--30~keV spectrum with a blurred relativistic reflection model in which the broad band spectrum is dominated by
reflection of coronal emission from an accretion disk.   This model was originally proposed to explain the spectral cutoff seen in the 7--10~keV band
as the result of relativistically smeared iron features \citetext{\citealp{caballero-garcia10}}.   Unlike the previously described models,
the reflection scenario predicts that above 10~keV the spectrum should flatten due to a contribution from Compton backscattering \citep{walton11}.  As expected, based on the sharp turnover persisting above 10~keV in the \nustar\ data, 
this model \citetext{the convolution of the {\tt reflionx} table \citep{ross05} with a Laor profile \citep{laor91}} overpredicts the data above $\sim$10~keV
(see Figure~\ref{fig:fitres1two}).   Similar results were found in broadband modeling including \nustar\ data of NGC~1313 X-1 \citep{bachetti13}, Circinus ULX5 \citep{walton13} and Holmberg IX X-1 \citep{walton14}.

\subsection{\xtwo\ Model Fits}

\begin{figure}
\includegraphics[width=\columnwidth]{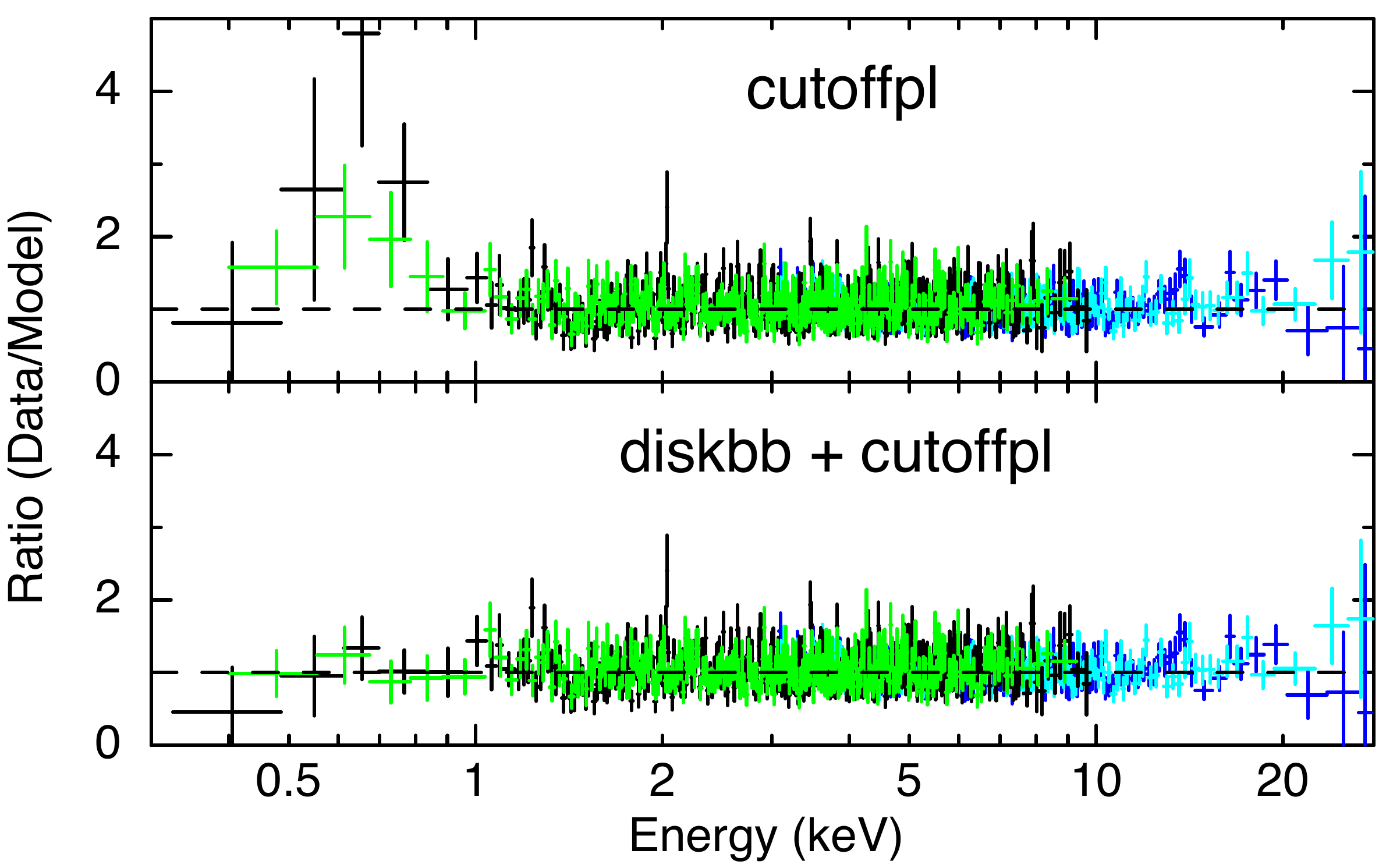}
\caption{Ratio of data to model for a cutoff power law (top), a cutoff power law with the addition of a
blackbody disk component (bottom) for \xtwo.  \xmm\ pn is plotted in black, MOS2
in green, \nustar\ FPMA in blue, and FPMB in light blue.  A clear excess can be seen at energies below $\sim$1~keV in top panel.}
\label{fig:fitres2one}
\end{figure}

\begin{figure}
\includegraphics[width=\columnwidth]{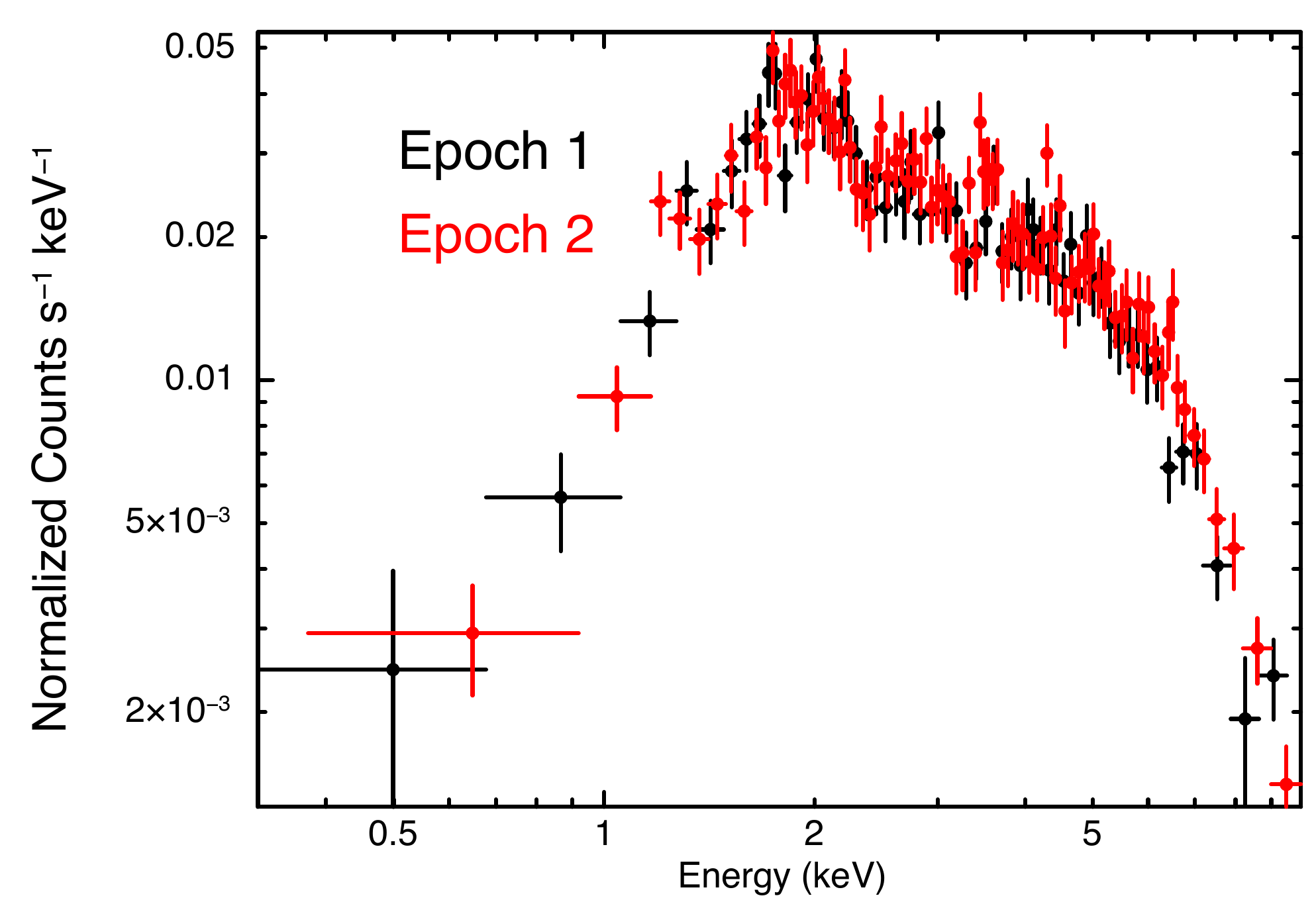}
\caption{Count spectrum of \xtwo\ for the two epochs of observations from \xmm\ pn. The spectra are  normalized to match 2--5~keV energy band. }
\label{fig:variabilityxtwo}
\end{figure}

We jointly fit the \xmm\ and \nustar\ spectra for the individual and combined epochs using an absorbed cutoff power law over the full 0.3--25~keV band.
The fit is acceptable, with $\chi^2$/dof of (1008/1039) for the combined data set, but clear residuals can be seen below 1~keV (see Figure~\ref{fig:fitres2one} top panel).   The cutoff power law 
parameters are very similar between epochs (see Table~\ref{tab:fitsxtwo_a}). 
Figure~\ref{fig:variabilityxtwo} directly compares the \xmm\ pn spectra between two epochs and shows that the spectra are consistent between epochs. 
If we add a disk blackbody component to fit the excess below 1~keV, 
the best-fit temperature is $T_{in} = 0.044^{+0.016}_{-0.007}$ keV (Figure~\ref{fig:fitres2one} bottom panel). Addition of a blackbody component improves the fit by $\Delta \chi^2$=18 for two additional degree of freedoms and statistically accounts for the excess, but with unrealistically high normalization for blackbody component (see Table~\ref{tab:fitsxtwo_a}).
The luminosity associated with the {\tt diskbb} component is extreme;  $4.5 \times 10^{43}$erg~s$^{-1}$ in the first epoch and $1.4 \times 10^{41}$erg~s$^{-1}$
in the second (in the 0.1--30.0 keV bandpass).  We conclude that the low energy excess is not associated with a blackbody component in the ULX \xtwo.  This unusual low-energy excess has been noted in previous \xmm\ observations of \xtwo. \citet{feng09} fit this component with a thermal plasma emission model down to 0.5~keV, but they note that the fit is poor below $\sim$0.5~keV, and cannot find a physical model that fits the data below $\sim$0.5 keV. A diffuse origin could be one viable explanation for this soft excess component. Therefore, we also tried fitting a collisionally ionized diffuse thermal plasma component (APEC model in XSPEC) along with the cutoff power-law model to the data. The diffuse component was only subject to a fixed Galactic absorption along the direction of IC 342 ($N_H$=3 $\times$ 10$^{21}$ cm$^{-2}$), while the cutoff power-law was still subject to an additional absorption column that was free to vary. This model also provides an acceptable fit with $\chi^2$/dof = 989/1037 and a best fit apec plasma temperature is 0.26$^{+0.15}_{-0.07}$. The cutoff power-law parameters are consistent with those listed in Table~\ref{tab:fitsxtwo_a}. The residuals look very similar to the bottom panel of Figure~\ref{fig:fitres2one}.

Alternatively, the soft X-ray excess could also result from incorrect modeling of absorber that might plausibly consists of material with non-solar abundances \citetext{see \citealp{goad06}, \citealp{pintore12}}. To test this possibility, we used $tbvarabs$ absorption model that allows us to vary the abundances for various elements. We refitted the absorbed cutoff power-law model with $tbvarabs$ by allowing abundance of various elements (C, N, O,  Ne, Na, Mg and Fe) to vary one-by-one. All the abundances were consistent with being solar, although the count statistics below ~1 keV is poor (see Figure~\ref{fig:fitres2one}), and it does not allow us to constrain any of the abundances. The soft excess does not get resolved and still remains. Hence, it is most likely that the soft X-ray excess is not due to non-solar absorbing material.  

\begin{figure}
\includegraphics[width=\columnwidth]{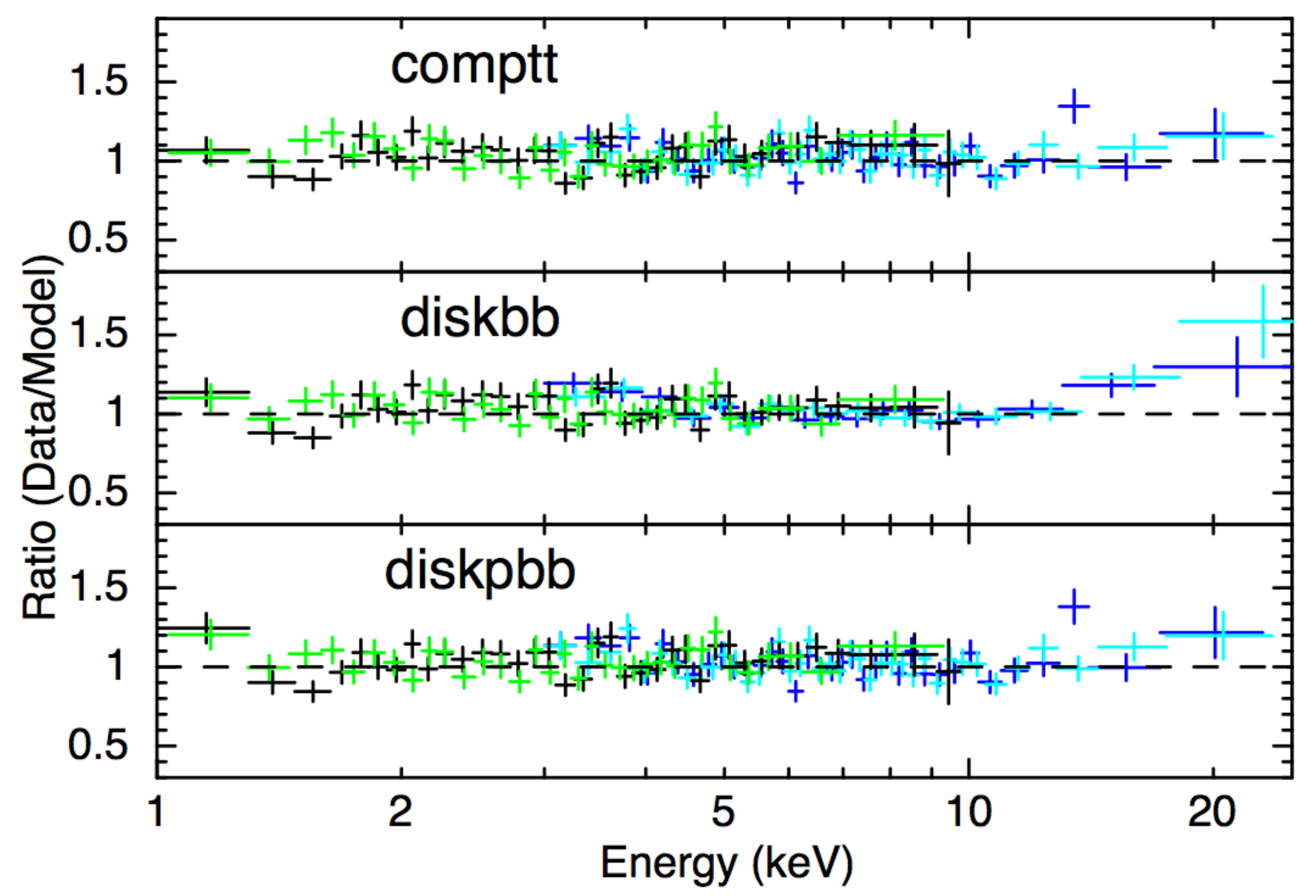}
\caption{Ratio of data to model for a Comptonization model (top), a blackbody disk (middle),  and an accretion disk with advection (bottom) for
\xtwo.   Data are fit in the 1--25~keV band.  \xmm\ pn is plotted in black, MOS2 in green, \nustar\ FPMA in blue, and FPMB in light blue.}
\label{fig:fitres2two}
\end{figure}

Since the modeling of the soft excess below $\sim$1 keV provides unrealistic results and its origin is not clear, we restrict fitting of additional models to the energy range 1--25~keV. 
Also, since the spectral shape above 1~keV is constant between the epochs we fit both epochs together to improve statistical accuracy given the lower count rate
of \xtwo\ relative to \xone.   We fit a Comptonization model with the seed photon temperature free to vary, since given the restricted energy range we cannot constrain the existence of a thin disk component.   This model provides a good fit to the data, with $\chi^2$/dof = 984/1017 (Table~\ref{tab:fitsxtwo_b}).   In contrast to \xone\, the disk blackbody model also fits the broadband data well, with $\chi^2$/dof = 1003/1019.  Above ~$\sim$10~keV the residuals are systematically slightly high (see Figure~\ref{fig:fitres2two}), although this is not significant.   If we allow the disk temperature profile to vary, the best-fit parametrization is $p = 0.68^{+0.03}_{-0.02}$, only slightly lower than for
a Shakura \& Sunyaev disk model. 

\begin{center}
\begin{deluxetable*}{ccccc}
\tablecolumns{5}
\tablewidth{0pc}
\tabletypesize{\scriptsize}
\tablecaption{Best fit spectral parameters for \xtwo\ from a joint fit to \xmm\
and \nustar\ data for a cutoff power-law model with and without a blackbody component. \label{tab:fitsxtwo_a}}
\tablehead{
\colhead{Parameter} & \colhead{Unit} & \colhead{Epoch 1} & \colhead{Epoch 2} &
\colhead{Combined Epochs} }
\startdata
\cutinhead{Model = TBabs $*$ cutoffpl} \\
$N_H$ & 10$^{22}$ cm$^{-2}$ & 1.85$\pm$0.21 & 1.67$\pm$0.20 & 1.78$\pm$0.14 \\
$\Gamma$ &  & 0.57$\pm$0.15 & 0.31$\pm$0.15 & 0.46$\pm$0.11 \\
$E_{cut}$ & keV & 5.8$^{+0.8}_{-0.6}$ & 5.0$^{+0.6}_{-0.5}$ & 5.4$^{+0.5}_{-0.4}$ \\
$N_{cpl}$ &  & (2.75$^{+0.48}_{-0.41}$)$\times$10$^{-4}$ & (1.95$^{+0.33}_{-0.28}$)$\times$10$^{-4}$ & (2.35$^{+0.28}_{-0.25}$)$\times$10$^{-4}$ \\
$\chi^2$/dof &  & 567/618 & 653/693 & 1008/1039 \\
Null hypothesis prob. &  & 0.93 & 0.86 & 0.75 \\
Flux & 10$^{-12}$ ergs cm$^{-2}$ s$^{-1}$ & 4.13$\pm$0.18 & 3.79$\pm$0.15 & 3.99$\pm$0.11 \\
Luminosity & 10$^{40}$ ergs s$^{-1}$ & 0.76$\pm$0.03 & 0.70$\pm$0.03 & 0.74$\pm$0.02 \\
\cutinhead{Model = TBabs $*$ (diskbb + cutoffpl)}
$N_H$ & 10$^{22}$ cm$^{-2}$ & 1.89$^{+0.20}_{-0.19}$ & 2.21$^{+0.50}_{-0.48}$ & 1.84$^{+0.15}_{-0.14}$ \\
$T_{in}$ & keV & 0.040$^{+0.019}_{-0.009}$ & 0.11$\pm$0.05 & 0.044$^{+0.016}_{-0.007}$ \\
$N_{bb}$ &  & 8.11$\times$ 10$^{9}$ & 1.01$\times$ 10$^{4}$ & 9.72$\times$10$^{8}$ \\
$\Gamma$ &  & 0.59$\pm$0.15 & 0.54$\pm$0.21 & 0.49$\pm$0.11 \\
$E_{cut}$ & keV & 5.9$^{+0.8}_{-0.6}$ & 5.7$^{+0.9}_{-1.0}$ & 5.5$^{+0.5}_{-0.4}$ \\
$N_{cpl}$ &  & (2.82$^{+0.49}_{-0.41}$)$\times$10$^{-4}$ & (2.51$^{+0.64}_{-0.54}$)$\times$10$^{-4}$ & (2.42$^{+0.29}_{-0.25}$)$\times$10$^{-4}$ \\
$\chi^2$/dof &  & 557/616 & 639/691 & 990/1037 \\
Null hypothesis prob. &  & 0.96 & 0.92 & 0.85 \\
Flux & 10$^{-12}$ ergs cm$^{-2}$ s$^{-1}$ & 4.16$^{+0.18}_{-0.17}$ & 3.99$^{+0.18}_{-0.17}$ & 4.03$\pm$0.12 \\
Luminosity & 10$^{40}$ ergs s$^{-1}$ & 0.77$\pm$0.03 & 0.74$\pm$0.03 & 0.74$\pm$0.02 \\
\enddata
\tablecomments{All the listed flux and luminosity estimates are unabsorbed values in 0.3--30 keV energy band.}
\end{deluxetable*}
\end{center}

\begin{deluxetable}{ccc}
\tablecolumns{3}
\tablewidth{0pc}
\tabletypesize{\scriptsize}
\tablecaption{Best fit spectral parameters for \xtwo\ from joint fit to \xmm\
and \nustar\ data with comptonization and disk blackbody models. \label{tab:fitsxtwo_b}}
\tablehead{
\colhead{Parameter} & \colhead{Unit} & \colhead{Combined Epochs} \\
\colhead{ } & \colhead{ } & \colhead{1+2} }
\startdata
\cutinhead{Model = TBabs $*$ comptt}
$N_H$ & 10$^{22}$ cm$^{-2}$ & 0.93$^{+0.17}_{-0.15}$ \\
T$_{0}$ & keV & 0.77$\pm$0.06 \\
kT$_{e}$ & keV & 3.24$^{+0.23}_{-0.19}$ \\
$\tau$ &  & 6.84$^{+0.59}_{-0.58}$ \\
$N_{comptt}$ &  & (1.85$\pm$0.15)$\times$10$^{-4}$ \\
$\chi^2$/dof &  & 984/1017 \\
Null hypothesis prob. &  & 0.77 \\
Flux & 10$^{-12}$ ergs cm$^{-2}$ s$^{-1}$ & 3.85$\pm$0.10 \\
Luminosity & 10$^{40}$ ergs s$^{-1}$ & 0.71$\pm$0.02 \\
\cutinhead{Model = TBabs $*$ diskbb}
$N_H$ & 10$^{22}$ cm$^{-2}$ & 1.75$\pm$0.07 \\
$T_{in}$ & keV & 3.53$\pm$0.08 \\
$N_{bb}$ &  & (1.28$\pm$0.12)$\times$10$^{-3}$ \\
$\chi^2$/dof &  & 1003/1019 \\
Null hypothesis prob. &  & 0.64 \\
Flux & 10$^{-12}$ ergs cm$^{-2}$ s$^{-1}$ & 3.84$\pm$0.08 \\
Luminosity & 10$^{40}$ ergs s$^{-1}$ & 0.70$\pm$0.02 \\
\cutinhead{Model = TBabs $*$ diskpbb}
$N_H$ & 10$^{22}$ cm$^{-2}$ & 2.05$^{+0.17}_{-0.16}$ \\
$T_{in}$ & keV & 3.82$^{+0.18}_{-0.17}$ \\
p &  & 0.68$^{+0.03}_{-0.02}$ \\
$N_{bb}$ &  & (7.52$^{+2.30}_{-1.78}$)$\times$10$^{-4}$ \\
$\chi^2$/dof &  & 990/1018 \\
Null hypothesis prob. &  & 0.73 \\
Flux & 10$^{-12}$ ergs cm$^{-2}$ s$^{-1}$ & 4.02$\pm$0.12 \\
Luminosity & 10$^{40}$ ergs s$^{-1}$ & 0.74$\pm$0.02 \\
\enddata
\tablecomments{All the listed flux and luminosity estimates are unabsorbed values in 0.3--30 keV energy band.}
\end{deluxetable}

\section{Short Timescale Variability}
\label{sec:timing}

We ran a timing analysis to search for non-coherent variability on fast timescales. We started by producing light curves cleaned from all intervals with increased background activity (such as the passages through the South Atlantic Anomaly) and gaps due to Earth occultation. From each light curve, we obtained power density spectra (PDS), the normalized square modulus of the Fourier Transform, by averaging the PDSs obtained from contiguous good intervals of data. Following the procedure 
described in \citet{bachetti13}, we search the PDS for features of different spectral width.      
For \xmm, we only used EPIC-PN data. The maximum frequency we investigated was $\sim$13 Hz, since the time resolution of the EPIC-PN camera in full frame mode is 73.4 ms.
The minimum frequency searched
was the inverse of each continuous data segment analyzed. For \nustar\ this was limited to $\sim$0.3~mHz due to the Earth occultations 
occurring on a 90~min timescale, and for \xmm\ it was $\sim0.1$~mHz. The PDS of both \xone\ and \xtwo\ are almost featureless.   Using a Kolmogorov-Smirnov test calculated from the light curve at different bin times,
we do not find any significant variability nor significant detection of QPOs or low-frequency noise.    

We also tested for variability using the normalized excess variance test \citetext{\citealp{edelson90}; \citealp{vaughan03}}. The excess variance, $F_{var}$, is a measure
of the intrinsic root mean square variability of the source.    For both ULXs during both epochs $F_{var}$ is consistent with 0.

\section{Summary and Discussion}
\label{sec:discussion}

\xmm\ and \nustar\ observed \xone\ and \xtwo\  simultaneously during two epochs spaced approximately one week apart.   Neither source
exhibited significant luminosity or spectral variability between epochs,
and neither source showed significant short-term variability during either epoch.   The
unabsorbed 0.3--30.0 keV luminosities of both ULXs were $1.04^{+0.08}_{-0.06} \times 10^{40}$\lum\ for \xone, and $(0.74\pm0.02) \times 10^{40}$\lum\
for \xtwo, derived using their respective best-fit spectral models.    This places both objects in the regime where \citet{sutton13} characterize ULXs based on their
spectral shapes to be predominantly in an  ultraluminous state.

The broadband 0.3--30~keV spectra of the two sources have generally similar features:  low-energy 
absorption above the Galactic value, and an exponential cutoff above $\sim8$~keV for \xone\ and $\sim6$~keV for \xtwo.   The derived absorption column
is model dependent; for \xone\ using the disk blackbody plus cutoff power law model it is $1.02 \times 10^{22}$~cm$^{-2}$, and for \xtwo\ it is $1.84 \times 10^{22}$ ~cm$^{-2}$ using a cutoff power law.   This column may be associated either with the ULX systems or their immediate environment, or it may be intrisic to the galaxy IC~342.    The overall curved spectral shape that continues above 10~keV in both sources has
now been consistently seen in other broadband X-ray spectra taken with {\em NuSTAR} 
\citetext{\citealp{bachetti13, walton13, walton14}} for
sources with $L \simgreat 10^{40}$\lum.  There are, however, notable differences in the spectral components derived for the two sources using models commonly used to characterize
ULX X-ray spectra.

\subsection{\xone}
\label{sec:xonediscussion}

\xone\ is known to have significant (factor of 2--3) flux variability as well as spectral variations on year timescales.   \cite{kubota01} analyzed multiple {\em ASCA} observations, and found that in two epochs spaced roughly seven years apart the flux had decreased by a factor $\sim$3 and the spectrum changed from a soft spectrum
modeled with a multicolor disk with $T\sim 1.8$~keV to a harder state modeled with a power law with $\Gamma \sim 1.7$.   With improved data quality in four epochs from \xmm,  \cite{feng09} confirmed the high/soft  state at luminosities $L_x\sim1.4\times10^{40}$\lum and a low/hard state at $L_x\sim5\times10^{39}$\lum. 
In our observations \xone\ appears to have observed in an intermediate state as compared to two states reported by
\citet{feng09}.   An absorbed power law with a blackbody disk ({\tt diskbb} in XSPEC) with $kT$ varying from $\sim 0.13$~keV (high state) to $\sim 0.3$~keV (low state) provides a relatively good fit to four epochs of \xmm\  data \citep{feng09}, although some curvature right around 10~keV is noted by these authors. \xone\ was included in the \xmm\ sample studied by \cite{gladstone09}, and this analysis found a break energy at $6.7^{+0.7}_{-1.0}$ keV when characterizing the spectrum with a broken power law. A Comptonization model can also be fit to the power law component, although previous constraints on the physical parameters are generally poor
owing to the limited bandpass.

The broadband combined \nustar\ and \xmm\  spectrum of \xone\ shows a clear high-energy cutoff extending above 10~keV.  A reflection model in its simplest version, where the continuum arises from relativistically blurred disk reflection \citep{caballero-garcia10} with a power-law without any cutoff,
is clearly ruled out by the persistent high energy cutoff.  
The best-fit model for the combined epochs 
is a power law with an exponential cutoff 
with $E_{cut} = 7.6^{+0.8}_{-0.7}$ keV and a low-energy excess modeled as a blackbody disk with $kT = 0.31 \pm 0.03$~keV.   There is little spectral variability between
epochs, and the luminosity varies only by $\sim$10\%.  A single blackbody disk is a poor fit to the broadband spectrum, however, allowing the disk temperature profile to deviate from the prescription for a Shakura \& Sunyaev thin disk does fit the data ($\chi^2$/dof = 2464/2344) with $p = 0.530\pm0.004$, significantly flatter than for a thin disk. However,  there is systematic structure in the residuals (see middle
panel of  Fig.~\ref{fig:fitres1two}) suggesting that more complex models are necessary for a good characterization. Modeling the spectrum with a blackbody disk plus Comptonization for the high-energy continuum yields an acceptable fit.

While several models provide a good characterization of the spectrum, the physical interpretation of the components is not clear.
The cutoff powerlaw characterization of the continuum is purely phenomenological.  The low-energy (0.2~keV)  black body plus
comptonization can be interpreted as blackbody disk emission dominating at low energies, and emission from an optically thick ($\tau \sim 5$), cool ($kT\sim 3$~keV) electron region upscattering the disk photons prevalent at high energy.  \citet{miller13} point out that 
such a cool, thick corona is physically unrealistic, likely being so physically large that it would not remain bound to the black hole.    Instead, these authors suggest
the two components may arise from a ``patchy" multi-phase disk, such as might be expected in a high Eddington rate system.    This model, while not unique, is consistent with the spectral data presented here, and provides a plausible interpretation. We note that in the disk plus comptonization model,  there are positive residuals at high energies, indicating an excess above $\sim$15~keV relative to this model.
While the significance  is low, we note that similar high energy excesses have now been seen in the \nustar\ spectrum of Circinus ULX5 \citep{walton13}, in Holmberg IX X-1 \citep{walton14} and NGC~5204 X-1 (Mukherjee et al. in preparation). This component could be associated with a hot corona similar to those found in the very high state of Galactic black hole binaries \citet{remillard06}. 
It should also be noted that the lack of intrinsic variability and best fit spectral parameters of \xone\ suggest that the source falls in the hard ultraluminous regime as defined by \citet{sutton13} with super-Eddington accretion modes where a massive outflowing radiatively driven wind is suggested to occur. The 
presence of such a massive wind in the form of a funnel-like geometry around the central regions of the accretion flow
could be another viable explanation of the observed characteristics of \xone. 

\subsection{\xtwo}
\label{sec:xtwodiscussion}
{\em ASCA} observations found that \xtwo\ also exhibits spectral and flux variability on timescales of years 
\citep{kubota01}.  \cite{feng09} found an unusual soft excess at $<1$~keV not describable by any emission model with sensible physical parameters.   Ignoring the low-energy component, these authors find little spectral variability in four \xmm\ observations with flux differing by a factor $\sim4$, in contrast to the {\em ASCA} observations.   The \xmm\ observations in the 1--10~keV band can all be fit either with an absorbed power law or with a disk blackbody with $kT$ between $\sim$2 - 3~keV, the latter being a somewhat better fit.

In our \xmm\ and \nustar\ data we also observe the unusual excess below 1~keV initially reported by \citet{feng09}. Several viable possibilities have been evaluated. 
A disk blackbody component statistically accounts for the excess in these observations (see Fig.~\ref{fig:fitres2one}), however, the corresponding luminosity is unphysically high, so this excess does not seems to be associated with blackbody emission.
On the other hand, the lack of variability could indicate a diffuse origin, possibly associated with an ionized gas component energized by
the ULX or by star formation. A collisionally ionized diffuse component also provides statistically acceptable fit for the excess. 
The lack of absorption associated with this very soft component, contrasting with the significant absorption suggested by model fits to the $E>1$~keV data for the ULX system, may suggest that the absorption is associated with the binary, and that the low-energy emission is more extended.  We have also checked the possibility of the absorbing material having non-solar abundances (like for Holmberg II X-1 see \citealp{goad06} and for NGC1313 see \citealp{pintore12}). However, allowing the abundances of various elements to vary does not account for the excess. Hence,  the true nature and origin of this component is unclear at the moment. 

Like all ULXs observed by \nustar\ to-date, the broadband spectrum of this source also shows a clear high-energy cutoff extending to 25~keV.   Ignoring data below 1~keV in order to avoid the soft excess which is unlikely to be associated with the binary, we get good fits using a cutoff power law with $E_{cut} = 5.5^{+0.5}_{-0.4}$~keV, a blackbody disk with $kT_{in} = 3.53\pm0.08$~keV, and a Comptonization model with $kT_e = 3.24^{+0.23}_{-0.19}$~keV and an optically thick scattering region.   For \xtwo, fitting the broadband emission as a disk does not require a temperature profile that is substantially broadened relative to a thin disk.

The physical interpretation of the different model fits is again not unique.   In the context of the blackbody disk fit, the 3.53~keV temperature is unusually high
but not unprecedented when compared to Galactic binaries.  \citet{tomsick05} reported a disk temperature of 3.2~keV for the black hole system
4U1630-47.   In the context of the Comptonization fit,  unlike for the \xone, there is no evidence in the spectrum for the seed photon source.   In our Comptonization 
model the seed photon temperature was left free to vary, and we find a best-fit value of $\sim$0.8~keV, hotter than that found for \xone.   For this temperature the seed photon emission should be visible even in the presence of the $\sim10^{22}$cm$^{-2}$ column.    If we force the seed temperature to be 0.3~keV, comparable to 
that found for \xone, we obtain a substantially worse fit ($\Delta \chi^2 \sim$ 60). It is possible that a lower-temperature (T$\sim 0.1$~keV) blackbody component is present, but is confused by  the combination of the E$<$~1~keV excess emission and the absorbing column.   The lack of evidence for a seed photon source makes the Comptonization interpretation 
less likely, and in this system the broad band emission may indeed arise from an optically thick, geometrically thin disk.

\section{Conclusions}
\label{sec:conclusions}

The first broadband (0.3--30 keV) X-ray observations of two ULXs in the galaxy IC~342 with \nustar\ and \xmm\ show that they have 
broadly similar spectra, with significant
local absorption and a clear exponential cutoff above $\sim$6--8~keV.  The two sources, \xone\ and \xtwo\ were caught in relatively bright states, with
luminosities of $1.04^{+0.08}_{-0.06} \times 10^{40}$\lum\ and $(0.74\pm0.02) \times 10^{40}$\lum\
respectively.   While a high energy cutoff has been suggested in previous
observations of \xone\ with \xmm, the instrumental bandpass was not sufficient to constrain the spectral shape, and we confirm it here with high significance.  
We can clearly rule out reflection from an accretion disc with relativistic line broadening surrounding an intermediate mass black hole \citep{caballero-garcia10}  as the origin of the X-ray emission.   This finding is consistent with the four other ULXs  with \nustar\ observations 
reported to-date \citetext{\citealp{bachetti13}, \citealp{walton13}, 2014},  which all have spectra distinct from classical states observed in Galactic black hole X-ray binaries.

Both objects have similar absorbing columns of $\sim 10^{22}$~cm$^{-2}$, significantly in excess of the Galactic value.   These columns are unusually high for ULXs, and may be either intrinsic to the galaxy, or associated with the binary systems.    The similarity of the columns might suggest an origin in the galaxy IC~342, however the column is high for a face-on spiral if  the objects are not embedded in a star forming region.   In \xtwo, we find an unabsorbed excess emission component at E$<1$~keV that could be associated with diffuse emission in the ULX environment.   This would support the idea that the absorption is intrinsic to the binary. If associated with the ULX systems the absorbing columns could indicate an inclined orientation with absorption due to a disk edge or wind.  \xone\ however has a hard spectrum, consistent with the hard ultraluminous state classification of \citet{sutton13}. Within this framework \xone\ would be a face-on system, at odds with the latter origin for the absorption.

The luminosity of \xone\ during these observations is only a factor $\sim$1.5 larger than that of \xtwo, and while the spectral curvature is broadly similar, fitting the
spectra to commonly used models shows clear differences.   Previous observations of \xone\ with \xmm\ alone are consistent with both power-law and Comptonization models for the high-energy continuum.   In the context of physical emission models we find that \xone\ is best described by a low-energy blackbody disk component plus a Comptonized continuum, although a small excess is also visible at the highest energies above 15~keV.   
As found for many other well-studied luminous ULXs the electron scattering region is cold and optically-thick.   This situation is physically problematic as the corona would be very extended \citep{miller13}.   An alternative model has the two components arising in a patchy, multiphase disk or a massive wind with funnel-like geometry, as might be expected in a high-Eddington rate system.   These scenarios are consistent with the broadband \xmm\ and \nustar\ observations.

If the \xone\ spectrum is to be associated with disk emission, the disk must have a significantly broader temperature profile compared to a standard thin disk.  Although there are clear residuals in the broadened disk model fit, the assumed p-disk profile is fairly simplistic and such a model is difficult to rule out.  In contrast, the
\xtwo\ spectrum can be fit with slight to no broadening of a thin-disk temperature profile.   Although the associated 3.8~keV disk temperature is higher than typically found in Galactic black hole binary disks, it is not unprecedented.
A Comptonization model with a cold, thick scattering medium also fits the \xtwo\ spectrum.   There is no evidence for a seed photon component in the
spectrum, and a free fit finds a seed blackbody temperature of 0.8~keV.   Fixing the seed temperature to the more typical value of 0.3~keV yields a poor fit.
The lack of evidence for the seed photon component favors the hot thin-disk interpretation, however we note that this evidence is indirect.

 In \xone\ we find some evidence for excess emission relative to the model fits and energies above
15~keV, although the significance is highly model-dependent.    Similar excesses, described by a power law of photon index $>$2, have been found in several other bright ULXs
observed with {\em NuSTAR} \citetext{see \citealp{walton13} for Circinus ULX5, \citealp{walton14} for Holmberg IX X-1 and Mukherjee et al. in preparation for NGC~5204 X-1}.  

For spectral components involving a temperature distribution, the relationship of temperature and luminosity can be used to constrain the origin.
Previous observations with \xmm\ of \xone\ show that in the {\tt diskbb + comptt} model, the low-energy thermal component does not vary as expected for
a thin disk \citep{feng09}, and this implies the origin is more likely due to a thick wind.  
Additional broadband observations with \nustar\ and \xmm\
of both ULXs in different luminosity states could better determine the relationship between the luminosity and temperature of the various spectral components, and
determine whether a disk origin is likely.  

Finally, for \xtwo, we confirm the unusual low-energy ($<$1~keV) spectral component found by \xmm\ \citep{feng09}, however, true origin of this component is not clear at present. 

\acknowledgments
This work was supported under NASA No. NNG08FD60C, and made use of data from the {\em Nuclear Spectroscopic Telescope Array} ({\em NuSTAR}) mission, a project led by Caltech, managed by the Jet Propulsion Laboratory, and funded by the National Aeronautics and Space Administration, and {\em XMM-Newton}, an 
ESA mission.  We would like to thank the anonymous referee for his positive comments to improve quality of this paper. We thank the {\em NuSTAR} Operations, Software and Calibration teams for support with the execution and analysis of these observations. This research has made use of the {\em NuSTAR} Data Analysis Software (NUSTARDAS) jointly developed by the ASI Science Data Center (ASDC, Italy) and Caltech (USA).
D. Barret and M. Bachetti are grateful to the Centre National d'Etudes Spatiales (CNES) for funding their activities.

\bibliography{ulx}

\end{document}